\documentclass[preprint]{JHEP3} 

\JHEPspecialurl{http://jhep.sissa.it/JOURNAL/JHEP3.tar.gz}

\usepackage{epsfig,multicol,mathbbol,amsmath,graphicx}

\newcommand\fverb{\setbox\pippobox=\hbox\bgroup\verb}
\newcommand\fverbdo{\egroup\medskip\noindent%
			\fbox{\unhbox\pippobox}\ }
\newcommand\fverbit{\egroup\item[\fbox{\unhbox\pippobox}]}
\newbox\pippobox
\newcommand{\be}{\begin{equation}}
\newcommand{\ee}{\end{equation}}
\newcommand{\bea}{\begin{eqnarray}}
\newcommand{\eea}{\end{eqnarray}}
\newcommand{\bml}{\begin{mathletters} \baselineskip 10pt}
\newcommand{\eml}{\baselineskip 12pt \end{mathletters}}

\newcommand{\mrm}{\mathrm}

\newcommand{\la}{\lambda}

\newcommand{\de}{\delta}

\newcommand{\bra}{\langle}
\newcommand{\ket}{\rangle}

\newcommand{\simleq}{\scriptstyle{\stackrel{<}{\sim}}}
\newcommand{\simgeq}{\scriptstyle{\stackrel{>}{\sim}}}

\newcommand{\pa}{\partial}

\newcommand{\D}{\mathcal{D}}

\newcommand{\fud}[2]{\frac{\delta #1}{\delta #2}}

\newcommand{\vcg}[1]{\mbox{\boldmath$#1$}}
\newcommand{\svcg}[1]{\mbox{\footnotesize\boldmath$#1$}}

\newcommand{\tr}{\mbox{tr}}

\title{Effective sigma models and lattice Ward identities}

\author{Leander Dittmann, Thomas Heinzl\thanks{Supported by DFG.}~
		and Andreas Wipf\\
	Theoretisch--Physikalisches Institut,
        Friedrich--Schiller--Universit\"at Jena, Max--Wien--Platz 1,
        07743 Jena, Germany\\
	E-mail: \email{l.dittmann@tpi.uni-jena.de},
        \email{t.heinzl@tpi.uni-jena.de}, \email{a.wipf@tpi.uni-jena.de}}

\preprint{FSU--TPI/05/02}           

\abstract{We perform a lattice analysis of the Faddeev--Niemi
effective action conjectured to describe the low--energy sector of
$SU(2)$ Yang--Mills theory. To this end we generate an ensemble of unit
vector fields (`color spins') $\vcg{n}$ from the Wilson action. The
ensemble does not show long--range order but exhibits 
a mass gap of the order of 1 GeV. From the distribution of color spins
we reconstruct approximate effective actions by means of  exact lattice
Schwinger--Dyson and Ward identities (`inverse Monte Carlo'). We show
that the generated ensemble cannot be recovered from a  
Faddeev--Niemi action, modified in a minimal way by adding an explicit
symmetry--breaking term to avoid the appearance of Goldstone modes.} 


\begin{document} 


\section{Introduction}

Confinement in pure Yang--Mills theory is still a theoretical
challenge. The problem actually has two faces. On the one hand, there is
confinement of static external sources in the fundamental representation
which manifests itself through the appearance of a linear potential
(nonzero string tension). On the other hand, there should be gluon
confinement implying a finite range of the gluonic interactions, i.e.\ a
mass gap. How the two faces are related is largely unclear at the
moment.

Recently, Faddeev and Niemi (FN) have suggested that the infrared
dynamics of glue might be described by the following low--energy
effective action \cite{faddeev:99a},
\be
\label{FN_ACTION}
  S_{\mathrm{FN}} = \int d^4 x \Big[ m^2 (\partial_\mu
  \vcg{n})^2 + \frac{1}{e^2} H_{\mu\nu}H^{\mu\nu} \Big] \; .
\ee
Here, $\vcg{n}$ is a unit vector field with values on $S^2$, $\vcg{n}^2
\equiv n^a n^a = 1$, $ a = 1,2,3$; $m$ is a dimensionful and $e$ a
dimensionless coupling constant.  The FN `field strength' is defined as
\be
  H_{\mu\nu} \equiv \vcg{n} \cdot \partial_\mu \vcg{n}\times
  \partial_\nu \vcg{n} \; .
\ee
Faddeev and Niemi argued that (\ref{FN_ACTION}) ``is the \textit{unique}
local and Lorentz--invariant action for the unit vector $\vcg{n}$ which
is at most quadratic in time derivatives so that it admits a Hamiltonian
interpretation and involves \textit{all} such terms that are either
relevant or marginal in the infrared limit''.

It has been shown that $S_{\mrm{FN}}$ supports string--like knot
solitons \cite{faddeev:97,battye:98,hietarinta:99}, characterized by
a topological charge which equals the Hopf index of the map
$\vcg{n}:S^3 \to S^2$. Here, $\vcg{n}$ is supposed to be static and
approaches a uniform limit at spatial infinity, $\vcg{n}_\infty =
\vcg{e}_z$. In analogy with the Skyrme model,
the $H^2$ term is needed for stabilization.  The knot solitons can
possibly be identified with closed gluonic flux tubes and are thus
conjectured to correspond to glueballs.  For a rewriting in terms of
curvature--free $SU(2)$ gauge fields and the corresponding
reinterpretation of $S_{{\rm FN}}$ we refer to \cite{vanbaal:01}.

In order for the model to really make sense, however, the following
problems have to be solved.  First of all, neither the interpretation
of $\vcg{n}$ nor its relation to Yang--Mills theory have been
fully clarified. An analytic derivation of the FN action requires
\begin{itemize}
\item
an appropriate change of variables, $A \to (\vcg{n}, X )$, relating the
Yang--Mills potential $A$ to $\vcg{n}$ and some remainder $X$
\item
the functional integration over $X$ to arrive at
an effective action $S_{\mathrm{eff}}$ for the $\vcg{n}$-field.
\end{itemize}
Some progress in this direction has been made
\cite{langmann:99,shabanov:99a,shabanov:99b,gies:01,freire:01,shabanov:01}
on the basis of the Manton--Cho decomposition \cite{manton:77,cho:80},
\be
\label{CHO}
  \vcg{A}_\mu = C_\mu \vcg{n} - \vcg{n} \times \partial_\mu \vcg{n} +
  \vcg{W}_\mu \; , 
\ee
where $C$ is an Abelian connection and $\vcg{n} \cdot \vcg{W}_\mu =
0$. Nevertheless, it is fair to say that there are no conclusive
results up to now.

Second, there is no reason why in a low--energy effective action for
the $\vcg{n}$--fields both operators in the FN `Skyrme term', which can
be rewritten as
\be
  H^2 = (\pa_{\mu}\vcg{n} \cdot \pa_{\mu}\vcg{n})^2 - (\pa_{\mu}\vcg{n}
  \cdot  \pa_{\nu}\vcg{n})^2 \; ,  
\ee
should have the same coupling. Third, and conceptually most important,
$S_{\mrm{FN}}$ has the same spontaneous symmetry breaking pattern as
the nonlinear $\sigma$--model, $SU(2)\to U(1)$. Hence, it should admit
two Goldstone bosons and one expects to find \textit{no} mass gap. 
In order to exclude these unwanted massless modes we have recently
suggested to break the global $SU(2)$ \textit{explicitly}
\cite{dittmann:01a}, an idea that has subsequently also been adopted by
Faddeev and Niemi \cite{faddeev:01}.

In what follows the FN hypothesis will be tested on the lattice. To
avoid the appearance of Goldstone bosons we allow for explicit 
symmetry--breaking terms.

\section{Generating an $\boldsymbol{SU(2)}$ lattice ensemble of
$\vcg{n}$--fields}

The conceptual problem to be solved in the first place is to obtain a
reasonable ensemble of $\vcg{n}$--fields. The (lattice version of the)
decomposition (\ref{CHO}) is of no help: it \textit{assumes} some
particular choice of $\vcg{n}$ on which the decomposition is
then based. One way of defining an $\vcg{n}$--field is via
Abelian gauge fixing, originally introduced by `t~Hooft
\cite{thooft:76a}. A prominent example in this class of gauges is the
maximally Abelian gauge (MAG) which is obtained via maximizing the
functional \cite{kronfeld:87a}
\be
\label{F_MAG} F_\mathrm{MAG}[U; g] \equiv \sum_{x,\mu} \tr \left( \tau_3
\, ^g U_{x,\mu} \tau_3 \, ^g U_{x,\mu}^\dagger \right) \equiv
\sum_{x,\mu} \tr \left(n_x U_{x,\mu} n_{x + \mu} U_{x,\mu}^\dagger
\right) \equiv \tilde{F}_\mathrm{MAG} [U; n] \; ,
\ee
with respect to the gauge transformation $g$. The maximizing $g$ then
defines the $\vcg{n}$--field according to
\bea
\label{NDEF} n_x \equiv g^\dag_x \, \tau_3 \, g_x \equiv \vcg{n}_x \cdot
\vcg{\tau} \; .
\eea
Instead of maximizing $F_\mathrm{MAG}$ with respect to $g$ one can
equivalently maximize $\tilde{F}_\mathrm{MAG}$ with respect to $n$
\cite{brower:97b} which results in the condition
\be
\label{LAPLACE} \triangle[U] \, n_x \equiv \lambda_x n_x \; .
\ee
Here, $\triangle[U]$ denotes the covariant Laplacian in the adjoint
representation (see App.\ \ref{APP:CONVENTIONS}), while $\lambda_x$ is a
Lagrange multiplier imposing that $\vcg{n}_x$ is normalized to unity,
locally at each lattice site $x$. In principle, (\ref{LAPLACE}) can be
solved for the field $\vcg{n}$ associated with the background
$U$. However, as this background is distributed randomly along its orbit
we will in turn obtain a random ensemble of $\vcg{n}$--fields
characterized by the two--point function
\be G_{xy}^{ab} \equiv \bra n^a_x n^b_y \ket = \frac{1}{3} \delta^{ab}
\delta_{xy} \; .
\ee
Thus, nontrivial correlations are absent. Fig.~\ref{FIG:RANDOM1} shows
that this is indeed what one gets in a typical Monte Carlo run.
\FIGURE[!ht]{\includegraphics[scale=0.6]{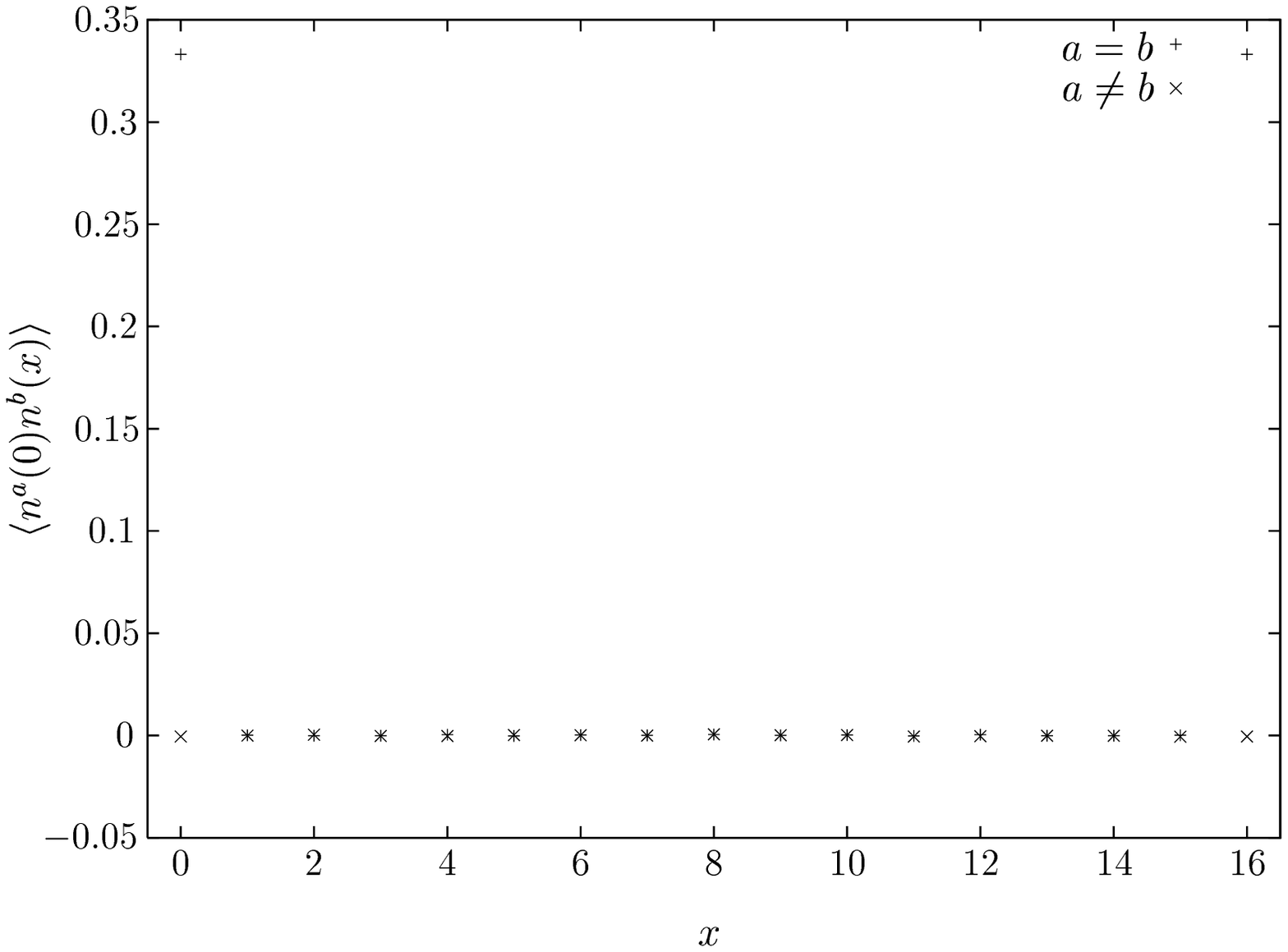}\caption{Behavior of
the two--point function $G_{xy}^{ab}$ along a lattice axis for a random
ensemble of $\vcg{n}$--fields, obtained via MAG and
(\protect\ref{NDEF}). Note that also the value 1/3 for $a=b$, $x=0$ and
$x=L$ is correctly reproduced.}
\label{FIG:RANDOM1}}
One way out of this problem is to follow the continuum approach of
\cite{shabanov:99b} which starts out with a covariant gauge fixing.
After having generated $SU(2)$ lattice configurations using the standard
Wilson action we therefore fix to lattice Landau gauge (LLG). The latter
is defined by maximizing the functional
\be
\label{F_LG} F_{\mathrm{LLG}}\equiv\sum_{x,\mu}\tr\,^\Omega U_{x, \mu},
\ee
with respect to the gauge transformation $\Omega$. In this way we
impose some `preconditioning' \cite{kovacs:99} which (i) eliminates the
randomness in our Yang--Mills ensemble and (ii) leaves a residual global
$SU(2)$--symmetry. The Landau gauge configurations are then plugged into
the MAG functional (\ref{F_MAG}) which subsequently is maximized with
respect to $g$. The gauge transformation $g$ obtained this way
determines $\vcg{n}$ according to (\ref{NDEF}). One may say that $g$
(and hence $\vcg{n}$) measure the gauge--invariant (!) distance between
the LLG and MAG gauge slices (see Fig.~\ref{FIG:LG_MAG}).
\FIGURE[ht]{\includegraphics[scale=0.4]{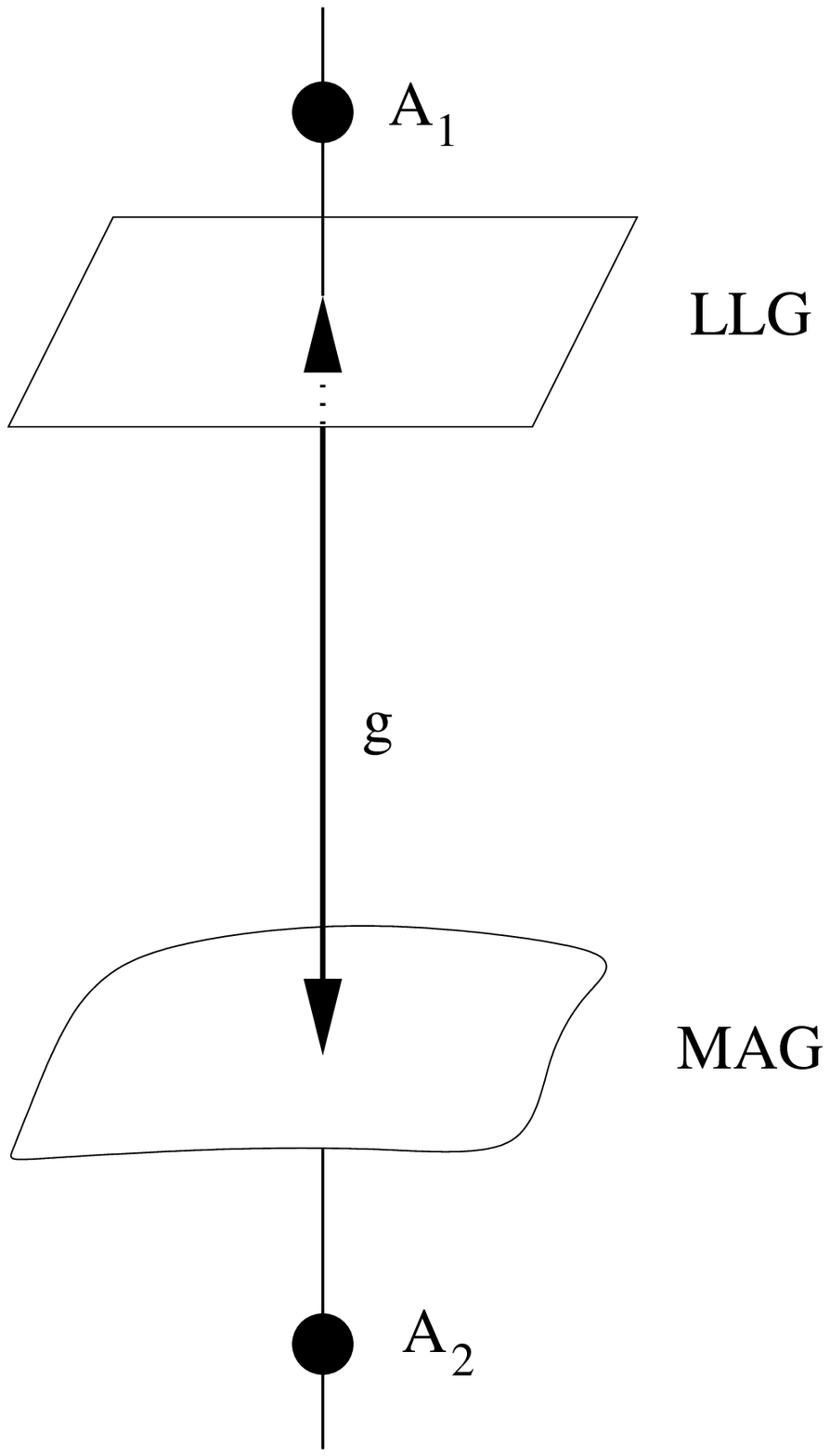}\caption{Gauge
invariant definition of $n \equiv g^\dagger \tau_3 g$. The gauge
equivalent configurations $A_1$ and $A_2$ are both mapped onto the same
`representatives' on the LLG or MAG slices (ignoring Gribov
copies). Thus, they are both associated with the \textit{same} gauge
transformation $g$ defining $n$.}\label{FIG:LG_MAG}}
In App.\ \ref{APP:LG_MAG} we show that LLG and MAG are `close' to
each other. Therefore, the maximizing $g$ is on average close to unity,
hence, on average, $\vcg{n}$ will be aligned in the positive
3--direction. In this way we have explicitly broken the global $SU(2)$
down to a global $U(1)$.

All computations have been done on a $16^4$--lattice with Wilson
coupling $\beta=2.35$, lattice spacing $0.13\ {\rm fm}$ and periodic
boundary conditions.  For the LLG we used Fourier accelerated steepest
descent \cite{davis:88} (see Fig.~\ref{FIG:LLGS}). The MAG was achieved
using two independent algorithms, one (AI) using iterations based on
elementary geometric manipulations (including overrelaxation steps), the
other (AII) being analogous to LLG fixing (see Fig.~\ref{FIG:MAGS}).
\FIGURE[]{\includegraphics[scale=0.6]{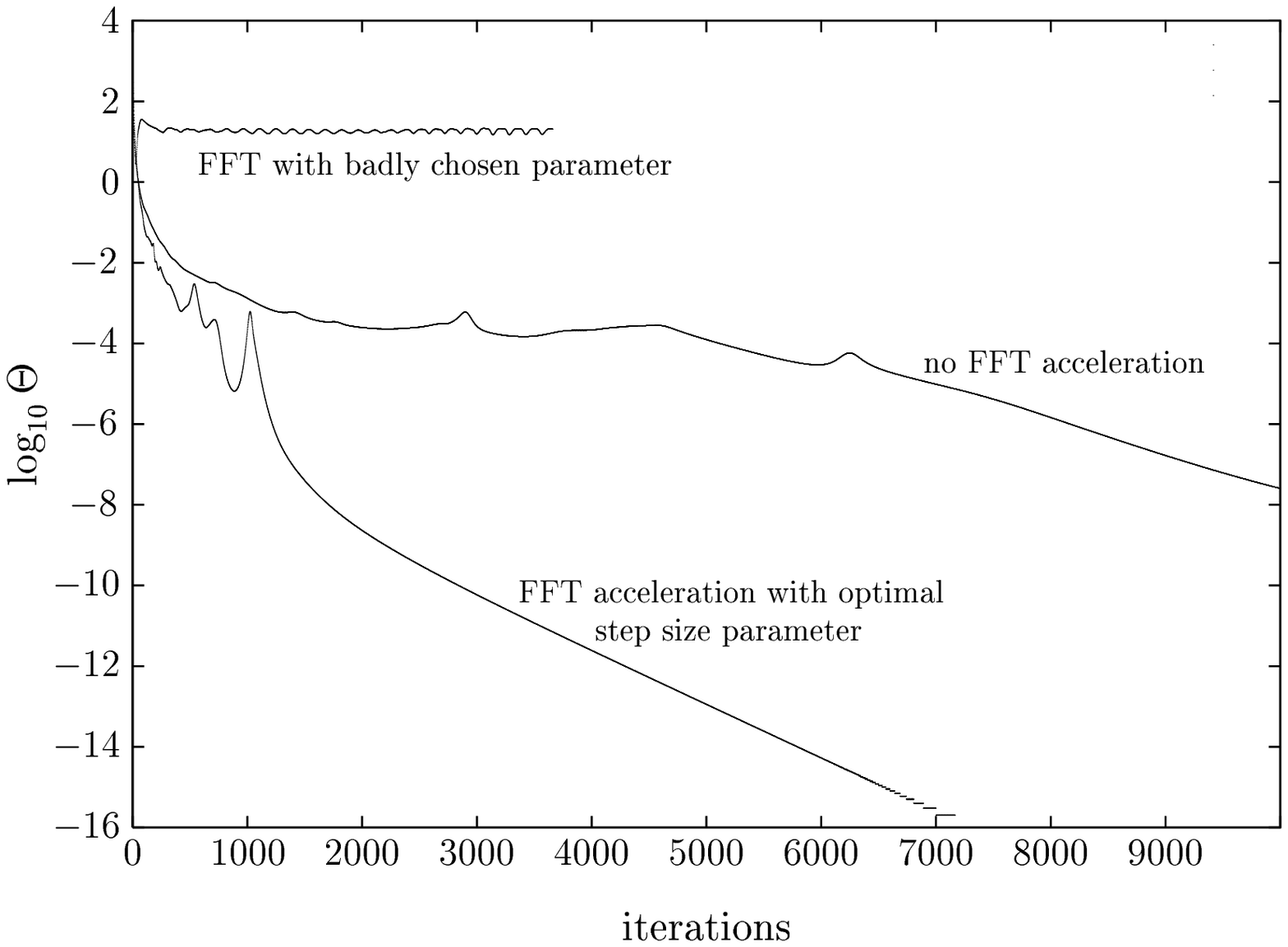} \caption{Behavior of
the LLG--functional using different algorithms. The parameter $\Theta$
measures the `distance' from the LLG, i.e. for $\Theta = 0$ the LLG is
achieved.}  \label{FIG:LLGS}}
\FIGURE[]{\includegraphics[scale=0.6]{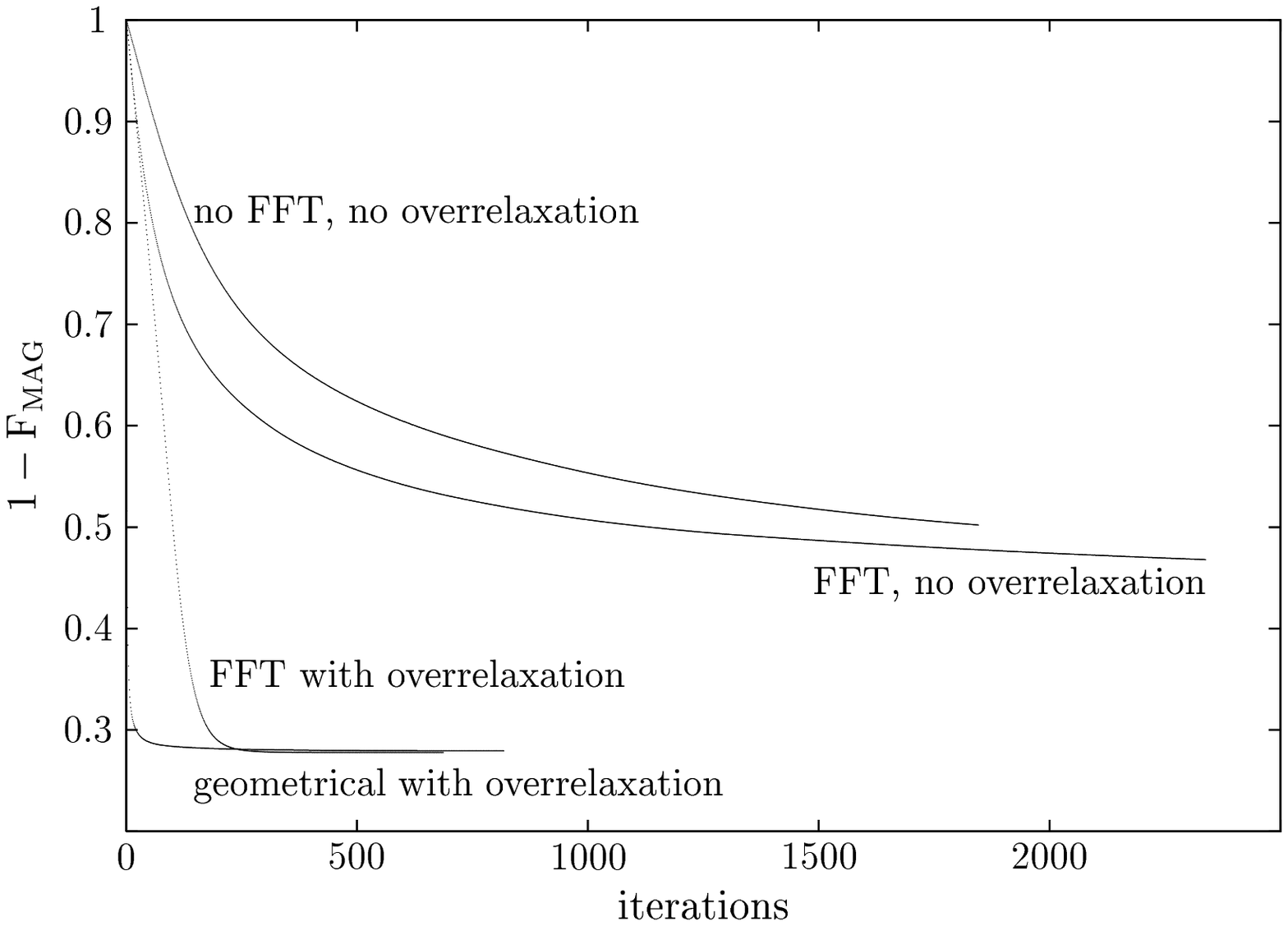} \caption{Behavior of
the MAG--functional using different algorithms.}  \label{FIG:MAGS}}

\section{Numerical results}

As expected, we observe a non--vanishing expectation value of the field
in the 3--direction, a 'magnetization' $\mathfrak{M}$ defined through $
\bra n^a \ket = \mathfrak{M} \, \de^{a3}$. Thus, the global symmetry is
indeed broken explicitely according to the pattern $SU(2)\to U(1)$. We
demonstrate this by exhibiting the angular distribution of the
$\vcg{n}$--field on its target space $S^2$ in
Fig.~\ref{FIG:ANG_DIST}. The azimuthal angle $\phi$ is equally
distributed, while the distribution of the polar angle $\theta$ has a
maximum near $\pi/2$ corresponding to the north pole, $\vcg{n} =
(0,0,1)$.
\FIGURE{\includegraphics[scale=0.6]{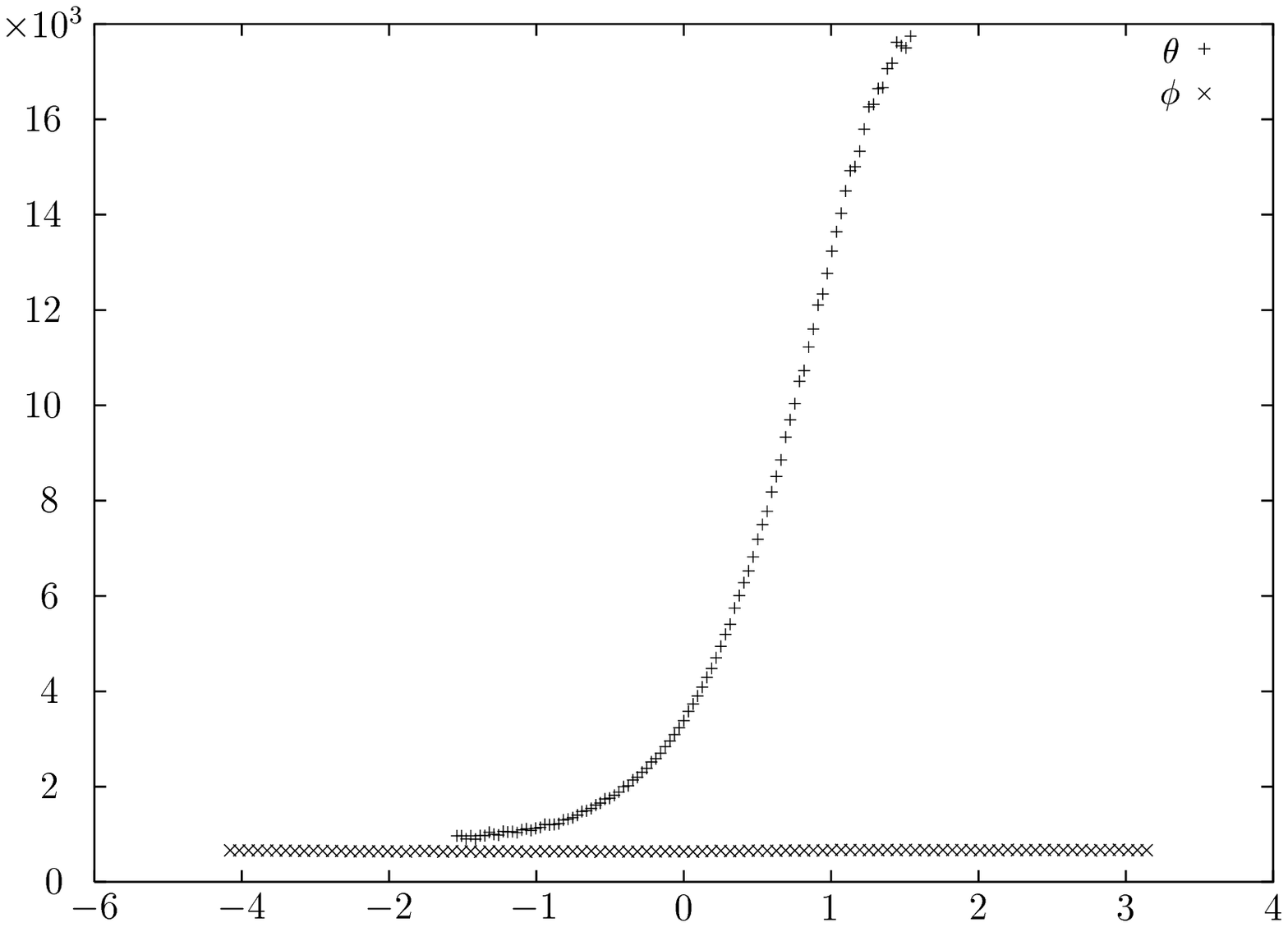}
\caption{Distribution of polar and azimuthal angles ($\theta$ and
$\phi$) associated with the unit vector $\vcg{n}$ on $S^2$. The uniform
distribution for $\phi$ and the maximum for $\theta = \pi/2$ shows that
$\vcg{n}$ is located near the north pole, $\vcg{n} =
(0,0,1)$.}\label{FIG:ANG_DIST}}

Explicit symmetry breaking also shows up in the behavior of the
two--point functions (Fig.~\ref{FIG:TWOPOINTFUNCTIONS}). The
longitudinal correlator, $G_x^\parallel \equiv \bra n_x^3 n_0^3 \ket
\sim \bra n^3 \ket \bra n^3 \ket = \mathfrak{M}^2$, exhibits clustering
for large distances, the plateau being given by the magnetization
(squared).
\FIGURE[ht]{\includegraphics[scale=0.6]{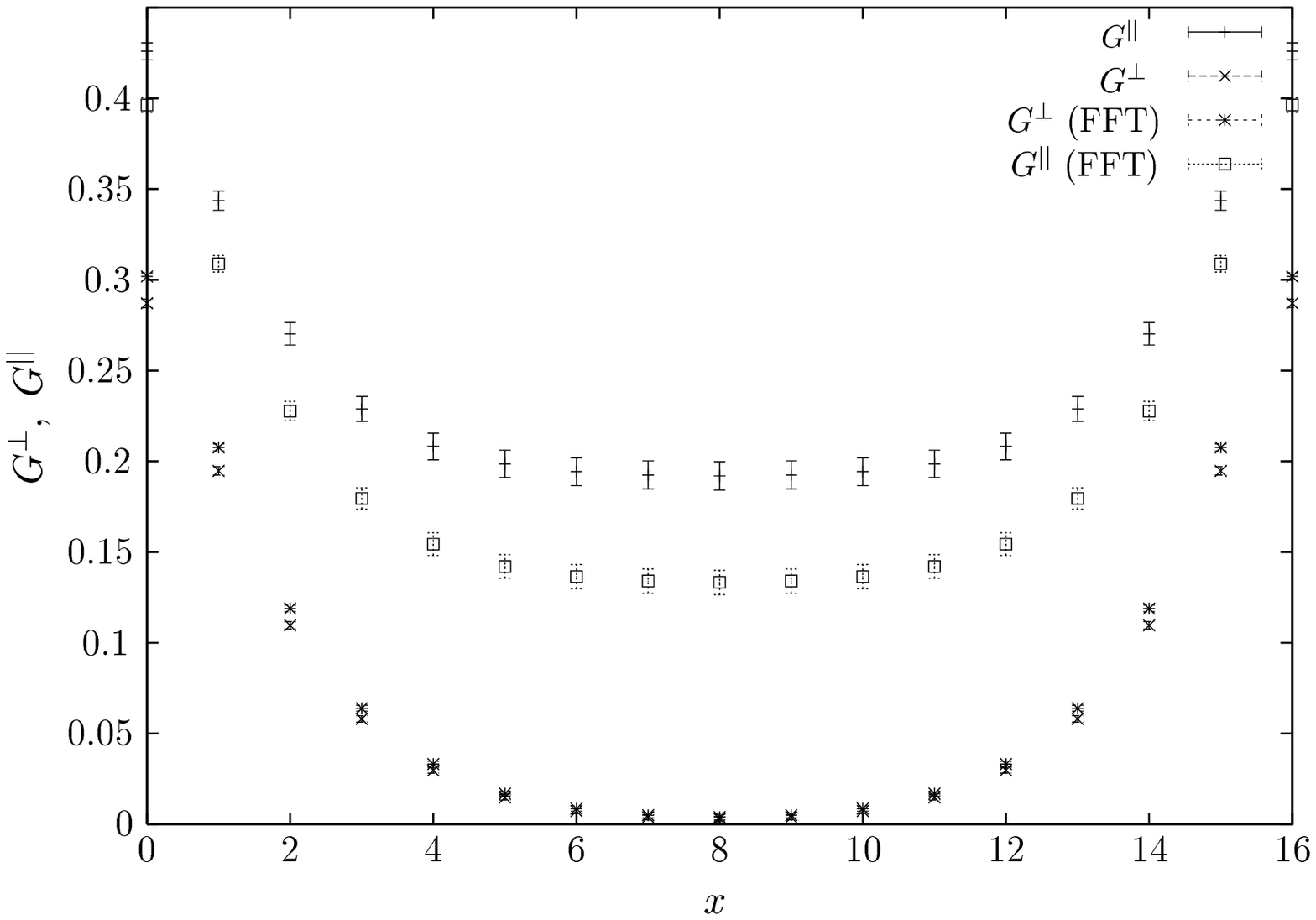}
\caption{Behavior of the two--point correlators of the $\vcg{n}$--field 
along a lattice axis (labelled by coordinate $x$). Note the difference
between algorithms AI and AII (FFT). Error bars 
exhibit the statistical error of the Monte Carlo simulation.}
\label{FIG:TWOPOINTFUNCTIONS}}
The transverse correlation function (of the would--be Goldstone
bosons)
\bea G_x^\perp \equiv G_{x0}^\perp \equiv \frac{1}{2} \sum_{i=1}^2 \bra
n_x^i n_0^i \ket \; ,
\eea
decays exponentially as shown in Fig.~\ref{FIG:COSHFITT}. This means
that there is a nonvanishing mass gap $M$ whose value can be obtained by
a fit to a $\cosh$--function (see Fig.~\ref{FIG:COSHFITT}).
\FIGURE[ht]{\includegraphics[scale=0.8]{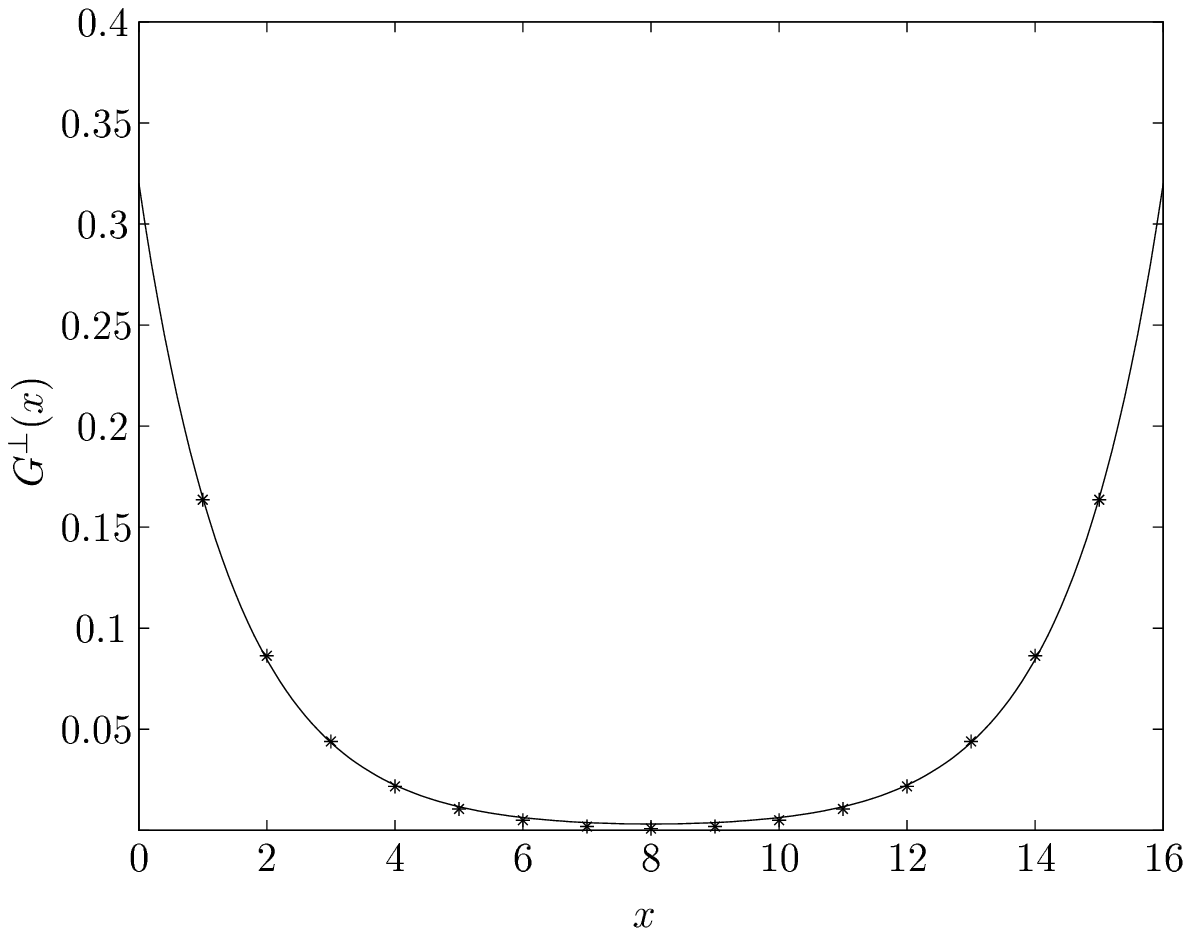}
\caption{The transverse correlation function along an arbitrary lattice
axis, fitted as $G^\perp (x) = a_1 \cosh(M(x - L/2)) + a_2$ with $a_1 =
0.0048$, $a_2 = -0.0053$, $M = 0.6084$. Data points are obtained with
algorithm AI.}
\label{FIG:COSHFITT}}
The numerical values of the observables, $\mathfrak{M}$, $M$ and the
transverse susceptibility,
\be
\label{TRANS_CHI}
  \chi^\perp \equiv \sum_x G_x^\perp \; ,
\ee
which all can be derived from the two--point functions, are summarized
in Table~1 for both algorithms.
\TABLE[ht]{\parbox{\textwidth}{\centering
\begin{tabular}{ccccc}
\hline algorithm & {$\mathfrak{M}$} & {$a^{-4}\chi^\perp$} & $aM$ & $M$
[GeV] \\
\hline AI & 0.438 & 92.57 & 0.61 & 0.95 \\ AII& 0.366 & 79.66 & 0.67 &
1.03 \\ \hline
\end{tabular}
\caption{Numerical results for some observables as obtained from the
longitudinal and transverse two--point functions, $G^\parallel$ and
$G^\perp$, respectively.}}}
The disagreement between AI and AII is statistically significant. We
attribute it to the ubiquitous Gribov problem \cite{gribov:78} (for
Abelian gauges, see \cite{bruckmann:00,bruckmann:01}). On the lattice,
this is the statement that maximizing gauge fixing functions like
$F_{\mathrm{MAG}}$ or $F_{\mathrm{LLG}}$ is equivalent to a spin--glass
problem with an enormous number of degenerate extrema. This implies that
the algorithms AI and AII will almost certainly end up in different
local maxima, which explains the difference between rows one and two in
Table~1.

As shown in the last column of Table~1, the numerical results for the
mass gap $M$ lead to a value of about 1 GeV in physical units.

$n^3$ is a local functional of the $n^i$, $n^3 = (1 - n^i
n^i)^{1/2}$. Thus, one expects the same exponential decay for the
longitudinal correlator $G^\parallel$. This can be confirmed with a
numerical value for the mass gap of $M = 0.66 \; a^{-1}$.

To improve statistics, we have calculated the time--slice correlator,
\be C^\perp (t) \equiv L^{-3} \sum_{\svcg{x}} G^\perp_{\svcg{x}, t} \; .
\ee
In the continuum, for purely exponential decay of $G^\perp$, this
would become proportional to a modified Bessel function $K_2$. An
associated fit works very well as is shown in Fig.~\ref{FIG:BESSELFIT}.
\FIGURE[ht]{\includegraphics[scale=0.8]{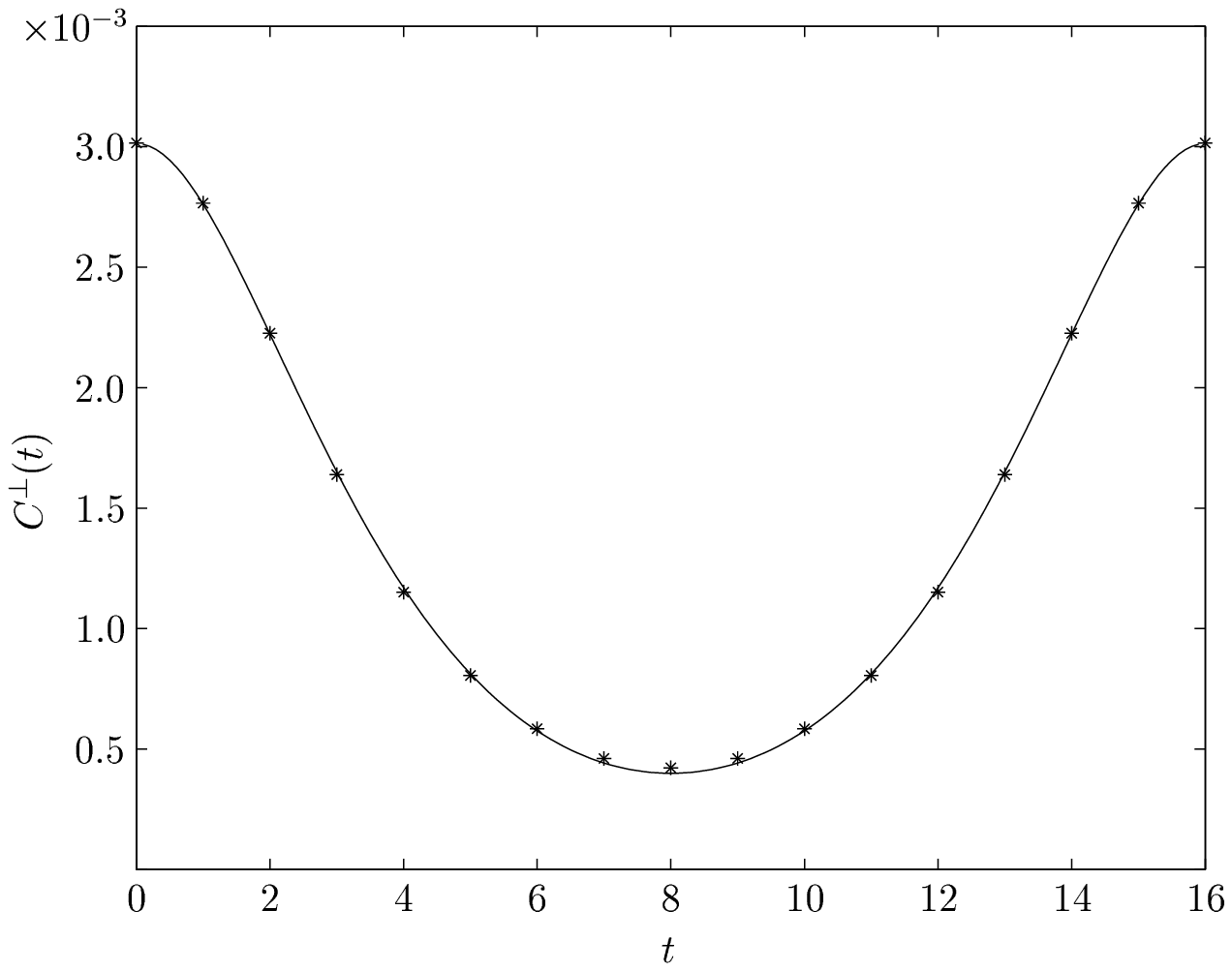}
\caption{The time--slice correlator fitted to a (properly symmetrized)
Bessel function, $C^\perp (t) = c_1 [t^2 K_2 (M t) + (t-L)^2 K_2 (M(L
-t))] + c_2$, where $c_1 = 0.0006$, $c_2 = 0.0001$ and $M =
0.6423$. Data points are obtained with algorithm AI.}
\label{FIG:BESSELFIT}}
Fitting the time--slice correlator according to
Fig.~\ref{FIG:BESSELFIT}, we obtain for the mass gap
\be aM = 0.642 \quad \mbox{i.e.} \quad M = 0.97 \; \mbox{GeV} \; .
\ee
This is the value with the smallest statistical errors.

The mass gap obtained differs significantly from the $SU(2)$ mass gap,
$M_{SU(2)} \simeq$ 1.5 GeV, obtained directly from a Wilson ensemble
with $\beta = 2.4$ \cite{teper:98}. We believe that the difference is
due to the highly nonlocal relation between the original Yang--Mills
degrees of freedom (the link variables) and the color spin
$\vcg{n}$. After all, we have implicitly solved the partial differential
equation (\ref{LAPLACE}) with link variables $U$ in Landau gauge
entering the adjoint Laplacian. The solution $\vcg{n}$ will clearly be a
nonlocal functional of these $U$'s. Consequently, we cannot expect that
the exponential decay of $G^\perp [n]$ will be governed by the lowest
excitation of the $U$--ensemble.

\section{Effective action and Schwinger--Dyson equations}

At this point it is natural to ask whether there is an effective action
$S_{\mathrm{eff}}[n]$ that reproduces the distribution of
$\vcg{n}$--fields leading to the results of the previous section.

At low energies, it should make sense to employ an ansatz in terms of a
derivative expansion,
\be
\label{ANSATZ} S_{\mathrm{eff}} = \sum_j \la_j S_j [\vcg{n}] +\sum_j
\la_j'\, S_j' [\vcg{n},\vcg{h}] \; , 
\ee
with $O(3)$ invariant operators $S_j$ and noninvariant operators
$S_j'$, which are ordered by increasing mass dimension.  Up to dimension
four, one has the symmetric terms,
\be
\label{SYMM} \begin{array}{lcl} S_1 = (\vcg{n} , \triangle \vcg{n}) \; ,
&& S_2 = (\vcg{n}, \triangle^2 \vcg{n}) \; , \\ S_3 = \big( \vcg{n}
\cdot \triangle \vcg{n}, \vcg{n} \cdot \triangle \vcg{n} \big) \; , &&
S_4 = \big( \vcg{n} \cdot \pa_\mu^\dagger \pa_\nu \vcg{n}, \vcg{n} \cdot
\pa_\mu^\dagger \pa_\nu \vcg{n} \big) \; , \end{array}
\ee
and the symmetry--breaking terms including a unit vector `source
field' $\vcg{h}$ \cite{dittmann:01a} (which can be thought of as the
direction of an external magnetic field),
\bea
\label{NONSYMM} S_1' = (\vcg{n} , \vcg{h}) \; , \quad S_2' = (\vcg{n}
\cdot \vcg{h}, \vcg{n} \cdot \vcg{h}) \; , \quad S_3' = (\vcg{n} \cdot
\triangle \vcg{n} , \vcg{n} \cdot \vcg{h}) \ .
\eea
In the above, we have introduced the scalar products
\be (f,g) \equiv \sum_x f_x g_x \; , \quad \vcg{u} \cdot \vcg{v} \equiv
u^a v^a \; ,
\ee
and the usual lattice Laplacian $\triangle$ (see
App.~\ref{APP:CONVENTIONS}).

Note that the $\vcg{n}$--field configurations are classified by the Hopf
invariant irrespective of the particular form of the (effective)
action. This, together with the usual scaling argu\-ments, shows that the
action (\ref{ANSATZ}) with the  operators (\ref{SYMM}) and
(\ref{NONSYMM}) should still support classical 
knot soliton solutions. Our ansatz thus does not exclude
this important feature.
 
The couplings in (\ref{ANSATZ}) can be determined by inverse Monte
Carlo techniques. The notion is suggestive: instead of creating an
ensemble from a given action, one wants to compute a (truncated)
action which gives rise to the given ensemble of $\vcg{n}$--fields. A
particular approach uses the 
Schwinger--Dyson equations \cite{falcioni:86,gonzalez-arroyo:87a}.
These represent an overdetermined linear system which can be used to
solve for the couplings in terms of correlation functions. The latter
are nothing but the coefficients of the linear system. 

For an unconstrained scalar field $\phi$, the Schwinger--Dyson
equations follow from translational invariance of the functional
measure, implying
\label{SD_FLAT}
\be
  0 = \int \D \phi \, \mathbb{P}_x \Big\{ F[\phi] \exp(-S[\phi])
  \Big\} \; ,
\ee
where $\mathbb{P}_x \equiv -i \delta/\delta\phi_x$ is the
(functional) momentum operator, and $F$ an arbitrary
functional\footnote{Usually one chooses $F [\phi] = \phi(x_1) \ldots
\phi(x_k)$.} of the field $\phi$. For a constrained field like
$\vcg{n}$ with a curved target space things are slightly more subtle
\cite{falcioni:86}. There is, however, a rather elegant way to derive
the Schwinger--Dyson equations if one exploits the isometries of the
target space $S^2$ \cite{gottlob:96}. The target space measure,
\be
\label{MEASURE}
  \D\vcg{n}=\prod_{x} d\vcg{n}_{x} \,\delta \big(\vcg{n}_x^2 - 1\big) \;
  , 
\ee
is obviously rotationally invariant, i.e.~under $\vcg{n}
\to R\vcg{n}$, $R\in O(3)$.  This implies the modified
Schwinger--Dyson identity 
\be
\label{SD_ANGMOM}
  \int \D\vcg{n} \, \vcg{L}_x \, \Big\{ F[\vcg{n}] \exp
  \big(-S_{\mathrm{eff}} [\vcg{n},\vcg{h}]\big) \Big\} = 0 \; ,
\ee
where $\vcg{L}_x$ denotes the angular momentum
operator (at lattice site $x$),
\be
  i \vcg{L}_x= \vcg{n}_x \times \frac{\partial}{\partial \vcg{n}_x}
  \quad \hbox{or} \quad
  iL_x^a \equiv \epsilon^{abc} n_x^b \fud{}{n_x^c}  \; . \ee
In shorthand--notation, (\ref{SD_ANGMOM}) can be rewritten as
\be
\label{SD_ANGMOM_SHORT}
  \bra \vcg{L}_x \, F[\vcg{n}] - F[\vcg{n}] 
  \vcg{L}_x\, S_{\mathrm{eff}}[\vcg{n}] \ket = 0 \;
  . 
\ee
These exact identities can be used to determine the unknown couplings 
$\la_j$. To this end one chooses a set of field monomials $F_i [\vcg{n}]$
and plugs them into (\ref{SD_ANGMOM_SHORT}) together with the
form (\ref{ANSATZ}) of the action. This yields the \textit{local} linear
system 
\be
\label{SD}
  \sum_j \bra F_i\, \vcg{L}_x \, S_j \ket \la_j
  +\sum_k \bra F_i\, \vcg{L}_x \, S_k' \ket \la_k' = \bra 
  \vcg{L}_x \, F_i \ket \; ,
\ee
which, in principle,  can be solved numerically, for instance by
least--square methods.  The identities obtained so far hold for
arbitrary actions $S_{\mathrm{eff}}[\vcg{n}]$. In particular, we have
not made use of any symmetries. Taking the latter into account will lead
to Ward identities.
 
Let us specialize to our lattice effective action (\ref{ANSATZ}). 
It is a sum of a symmetric part $S$ containing the terms
(\ref{SYMM}) and an asymmetric part $S'$ containing the terms
(\ref{NONSYMM}),
\be
  S_{\mathrm{eff}} = S[\vcg{n}] + S' [\vcg{n},\vcg{h}] \; .
\ee
Due to the invariance of $S$ under global $O(3)$ rotations it is an
$O(3)$--singlet and hence annihilated by the total angular momentum,
\be
  \vcg{L} S=0,\qquad \vcg{L}=\sum_x\vcg{L}_x \; ,
\ee
such that $\vcg{L} S_\mathrm{eff} = \vcg{L}S'$.  Thus, summing over
all lattice sites $x$ in (\ref{SD_ANGMOM_SHORT}) yields the (broken) 
Ward identity,
\be
\label{WARD}
  \bra \vcg{L}F[\vcg{n}] - F[\vcg{n}] \, \vcg{L}S'[\vcg{n},\vcg{h}]
  \ket=0 \; , 
\ee
where the second terms contains the infinitesimal change of the
non-invariant part $S'$ of the effective action under rotations of
$\vcg{n}$. Note that the coupling constants $\lambda_j$ of the
$O(3)$--symmetric operators $S_j$ have disappeared in the
Ward identity (\ref{WARD}) so that only the symmetry--breaking couplings
$\lambda'$ are present.  We have collected the explicit lattice 
Schwinger--Dyson  and Ward identities used in our simulations
in App.\ \ref{APP:WARD}. As the former are local relations, they
naturally contain more information than the global Ward identities. In
particular, one does have access to length scales.

\section{Comparing Yang--Mills and FN ensembles}

\subsection{Leading--order ansatz}

To leading order (LO) in the derivative expansion we have a standard
nonlinear sigma model with symmetry--breaking term, 
\be
\label{S_LO}
  S_\mathrm{eff} = \sum_x (\lambda \vcg{n}_x \cdot \triangle
  \vcg{n}_x + \lambda^\prime \vcg{n}_x \cdot \vcg{h}) \; , \quad \vcg{h}
  \equiv \vcg{e}_z \; .
\ee
Inverse Monte Carlo amounts to determining the couplings $\lambda$ and
$\lambda'$ such that the probability distribution associated with the LO
action (\ref{S_LO}) fits the observables of the Yang--Mills ensemble of
$\vcg{n}$--fields\footnote{Throughout this section, we
refer to algorithm AI.}. The associated Schwinger--Dyson
equation (\ref{SD}), with $F[\vcg{n}] = n_x^a$, can be written as  
\be
\label{SDE_LO}
  \lambda H_{xy}  + \lambda' G^\perp_{xy} = - \mathfrak{M} \,
  \delta_{xy} \; ,  
\ee
where $H$ denotes the (antisymmetrized) two--point function of
$n^i$ and $n^i \triangle n^3$,
\be
  H_{xy} \equiv \bra n_x^i n_y^i \triangle n_y^3 \ket - \bra n_x^i n_y^3
  \triangle n_y^i \ket \equiv \bra n_x^i n_y^{[\, i} \triangle n_y^{3]}
  \ket  \; . 
\ee
To analyse (\ref{SDE_LO}) we define a `reduced' two--point function
$h_{xy}$ and magnetization $\mu$,
\be
  h_{xy} \equiv H_{xy}/G_{xy}^\perp \; , \quad \mu \equiv
  \mathfrak{M}/G_{xx}^\perp \; , 
\ee
and rewrite (\ref{SDE_LO}) as the inhomogeneous system (using
translational invariance to replace $x-y \to x$),
\bea
  \lambda h_x + \lambda' &=& 0 \; , \quad x = 1, \ldots , 8 \; ,
  \label{LO_SYSTEM1} \\
  \lambda h_0 + \lambda' &=& - \mu \; . \label{LO_SYSTEM2}
\eea 
The solution is found to be
\bea
  \lambda &=& \phantom{-} \frac{1}{h_x - h_0} \, \mu \; ,
  \label{LO_SOL1} \\ 
  \lambda' &=& - \frac{h_x}{h_x - h_0} \, \mu  \; ,
  \label{LO_SOL2} 
\eea
with the numerical boundary value given by $h_0 = 0.1410$ (cf.\ 
Fig.~\ref{FIG:H_YM}).  
Clearly, the system (\ref{LO_SYSTEM1}), (\ref{LO_SYSTEM2}) is
overdetermined (nine equations for two unknowns). This is reflected in
the fact that $\lambda$ and $\lambda'$ in (\ref{LO_SOL1}) and
(\ref{LO_SOL2}) depend on the lattice \textit{distance}
$x$ via $h_x$. If the Yang--Mills ensemble were exactly
described by the LO action (\ref{S_LO}), there would be no such 
$x$--dependence. Rather, for any $x = 1, \ldots , 8$, we would have the
same values for $\lambda$ and $\lambda'$, respectively. Thus, to test
the quality of the LO ansatz, we divide (\ref{LO_SOL2}) by
(\ref{LO_SOL1}) showing explicitly that $h_x$ should be constant, 
\be
\label{H_LO}
  h_x = -\frac{\lambda'}{\lambda} \equiv - \kappa' = const. \;
  , \quad x  \ne 0 \; . 
\ee
Fig.~\ref{FIG:H_YM} shows that this is not the case.
\FIGURE[ht]{\epsfig{file=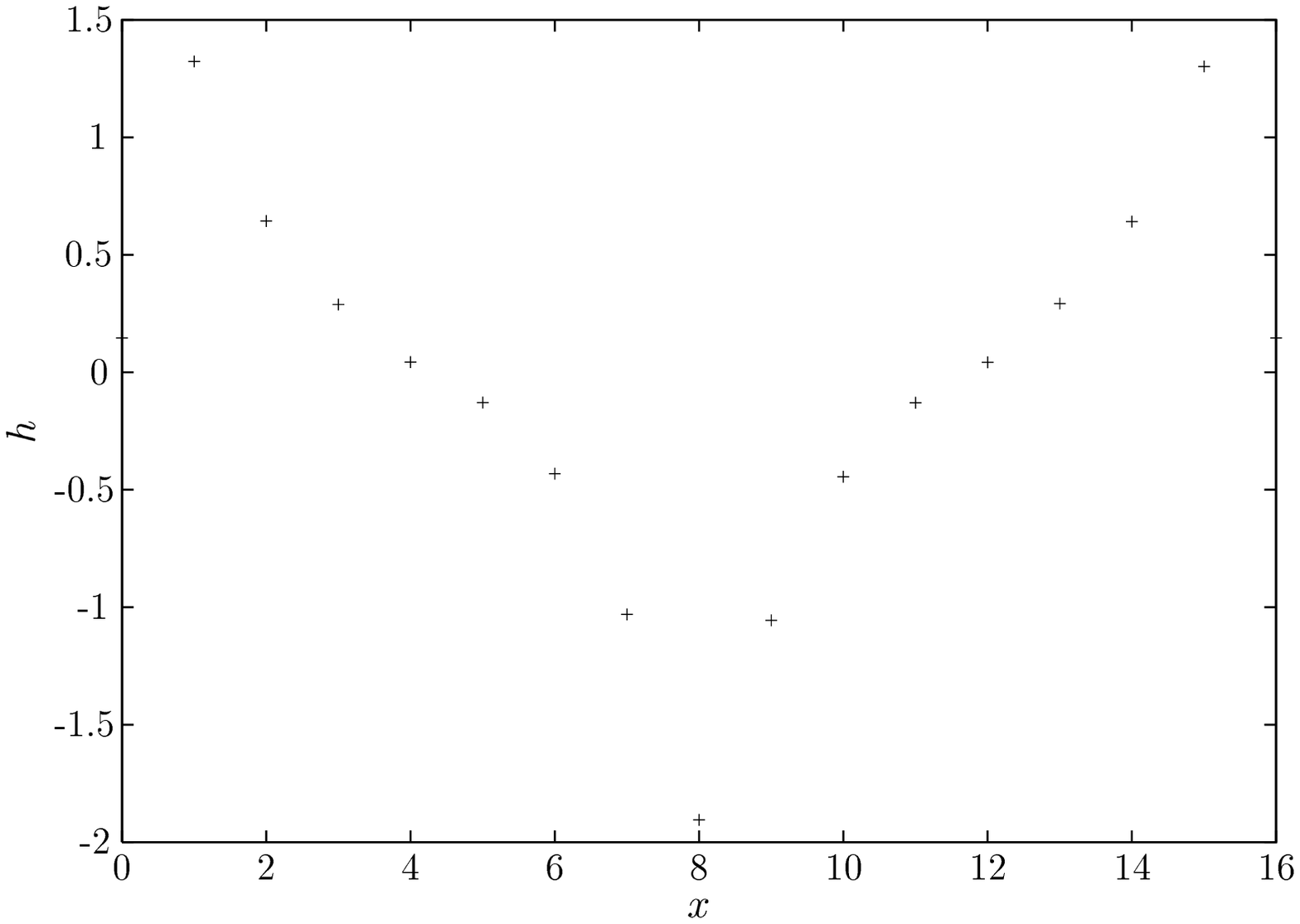,scale=0.6}
\caption{$h_x$ for the Yang--Mills ensemble. The boundary value is $h_0
= 0.1410$.} 
\label{FIG:H_YM}}
Therefore, a minimal sigma model with symmetry--breaking term does not
yield a good representation of our Yang--Mills ensemble of
$\vcg{n}$--fields.  If we nevertheless insist on the LO description, we
have to `fit' $h_x$ by a horizontal line so that the numerical
determination of the couplings via (\ref{LO_SOL1}) and (\ref{LO_SOL2})
is beset by large errors, 
\bea
  \lambda &=& -1.41 \pm 5.25 \; , \label{L_NUM_LO}\\
  \lambda' &=& -1.33 \pm 0.74 \; . \label{LP_NUM_LO}
\eea
Obviously, $\lambda$ (including its sign) remains essentially
undetermined. For $\lambda'$ the situation is slightly better.  

In order to assess the errors it is worthwhile to check whether our
numerical accuracy is sufficient to really validate the
Schwinger--Dyson identity (\ref{SDE_LO}) for the LO action (\ref{S_LO}) 
on the lattice. To this end we have simulated
(\ref{S_LO}) with a combination of Metropolis and cluster algorithms
producing a number of 150 configurations using the central values
(\ref{L_NUM_LO}) and (\ref{LP_NUM_LO}) as the input couplings. The
result for $h_x$ in the LO ensemble is presented in Fig.~\ref{FIG:H_LO}.
\FIGURE[ht]{\epsfig{file=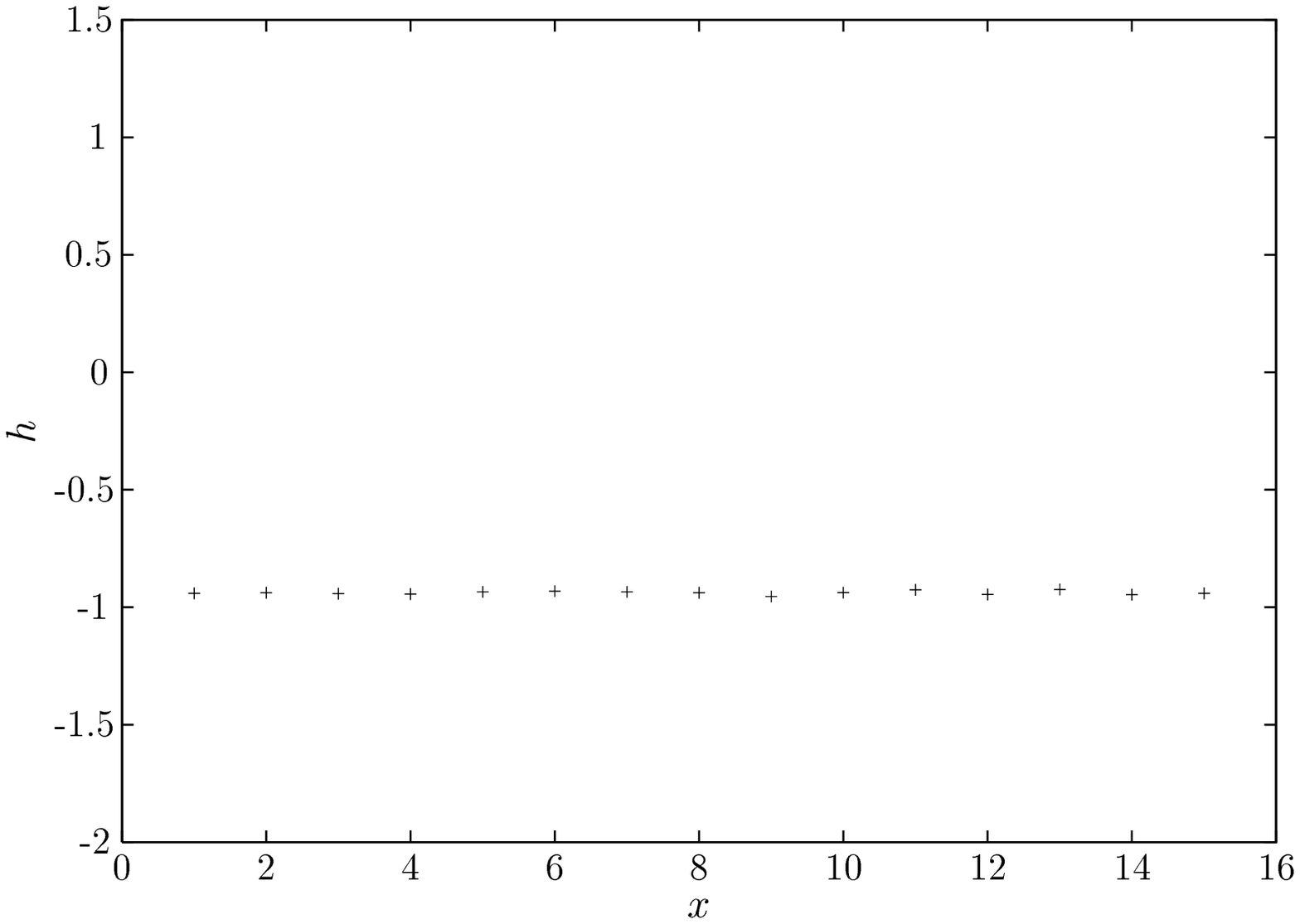,scale=0.6}
\caption{$h_x$ for the LO ensemble. Fitting the data points yields $h_x
= - \kappa' = -0.938 \pm 0.002$.}
\label{FIG:H_LO}}
It is reassuring to note that the simulation of the minimal sigma
model reproduces the input value $\kappa' = 0.943$ very well (for $x \ne
0$), the error being of the order of one percent. The prediction
(\ref{H_LO}) thus can be verified with high accuracy for the LO action
(\ref{S_LO}). We conclude that inverse Monte Carlo works quite well when
applied to the minimal $\sigma$--model.

The discrepancy between the LO and Yang--Mills ensembles can be 
further visualized by looking at the susceptibility. For the action
(\ref{S_LO}) and the choice $F[\vcg{n}] = n_x^a$, the Ward identity
(\ref{WARD}) assumes the simple form 
\be
\label{WARD_LO}
  \chi^\perp = -\mathfrak{M}/ \lambda' \; .
\ee
A consistency check is provided by noting that this can directly be
obtained by summing (\ref{SDE_LO}) over $x$. Plugging in the
magnetization from Table~1 and $\lambda'$ from (\ref{LP_NUM_LO}) we find
\be
  \chi^\perp = 0.33 \pm 0.18 \; ,
\ee
This is way off the Yang--Mills value of 92.6 displayed in Table~1. For
magnetization and mass gap the simulation of the LO ensemble yields the
values 
\be
  \mathfrak{M} = 0.93 \; , \quad M = 1.5 \; , 
\ee
which are both larger than the Yang--Mills values of Table~1.

The discussion of this subsection thus shows quite clearly that more
operators will have to be included in order to possibly make inverse
Monte Carlo work reasonably well.

\subsection{FN action with symmetry--breaking term}
        
In this subsection we consider the FN action (\ref{FN_ACTION}) with a LO
symmetry--breaking term,
\be
\label{S_NLO}
  S_\mathrm{eff} = \sum_x \left\{ \lambda \vcg{n}_x \cdot
  \triangle \vcg{n}_x + \lambda_{\mbox{\tiny FN}} \left[ (\vcg{n} \cdot
  \triangle \vcg{n})^2 - (\vcg{n} \cdot \pa_\mu^\dagger \pa_\nu \vcg{n})^2
  \right] + \lambda^\prime \vcg{n}_x \cdot \vcg{h} \right\} \; .
\ee
This ansatz does not include all terms of next--to--leading order (NLO)
in the derivative expansion. It should be viewed as a minimal
modification of the original FN action by adding an  explicit
symmetry--breaking term to obtain a mass gap.

The Schwinger--Dyson equation generalizing (\ref{SDE_LO}) becomes
\be
\label{SDE_NLO} \lambda H_{xy} + \lambda_{\mbox{\tiny
FN}}H_{xy}^{\mbox{\tiny FN}} + \lambda' G_{xy}^\perp = -\mathfrak{M} \,
\delta_{xy} \; .
\ee
The new two--point function $H^{\mbox{\tiny FN}}$ is given by
(\ref{H^FN}). The local identities (\ref{SDE_NLO}) are to be solved for
the three unknown couplings $\lambda$, $\lambda'$ and
$\lambda_{\mbox{\tiny FN}}$.
\FIGURE[ht]{\epsfig{file=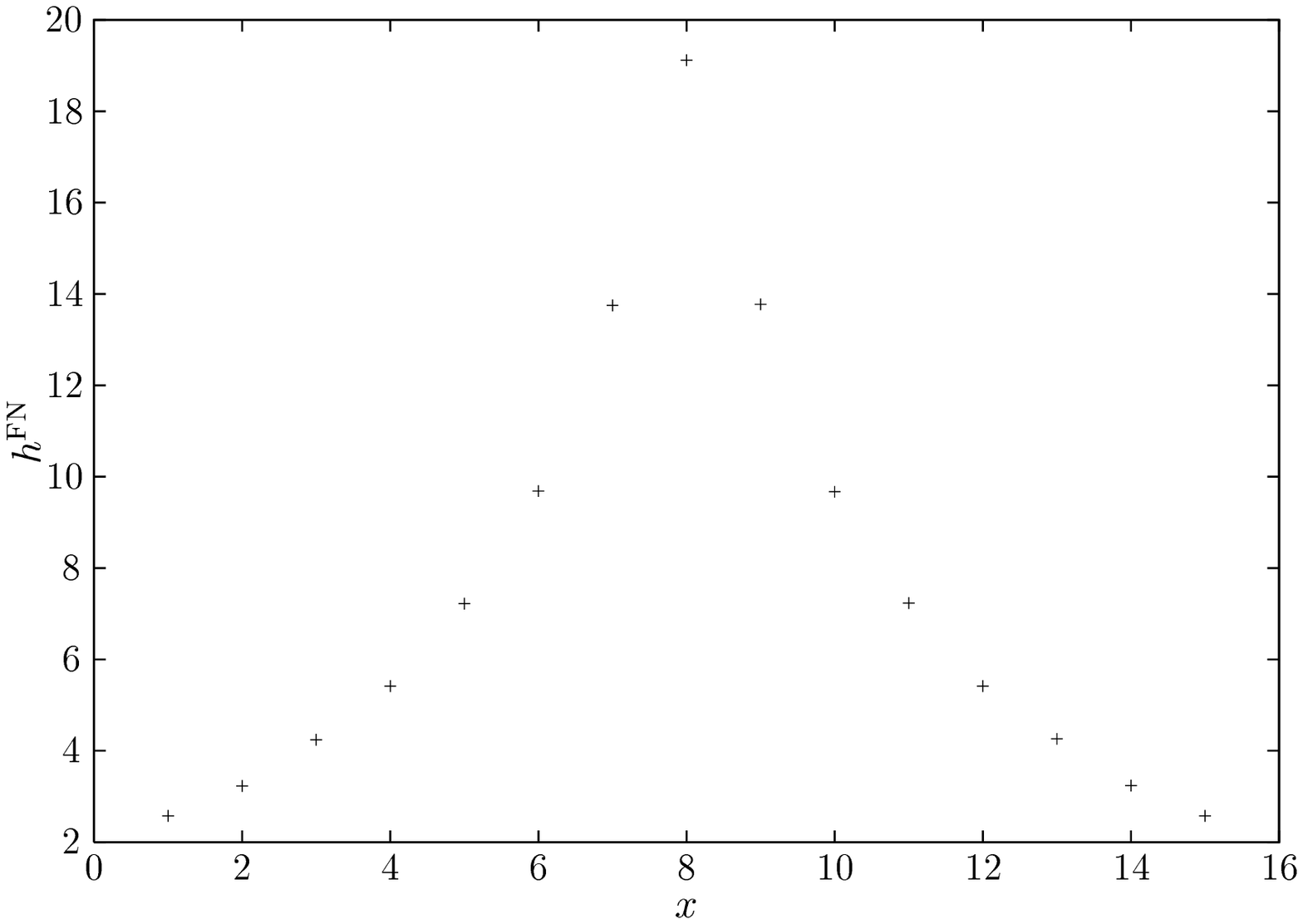,scale=0.6} 
\caption{$h^{\mbox{\tiny FN}}_x$ for the Yang--Mills
ensemble. The boundary value $h_0^{\mbox{\tiny FN}} = -31.24$ is not
displayed.}
\label{FIG:H_FN_YM}}
Introducing another reduced two--point function, 
\be
  h_{xy}^{\mbox{\tiny FN}} \equiv H_{xy}^{\mbox{\tiny FN}}/G_{xy}^\perp 
  \; ,
\ee
which is plotted in Fig.~\ref{FIG:H_FN_YM},  we obtain, instead of
(\ref{LO_SYSTEM1}) and (\ref{LO_SYSTEM2}), the (overdetermined) system,  
\bea
  \lambda h_x + \lambda_{\mbox{\tiny FN}} h_x^{\mbox{\tiny FN}} +
  \lambda' &=& 0 \; , \quad x = 1, \ldots , 8 \; ; \label{NLO_SYSTEM1} \\
  \lambda h_y + \lambda_{\mbox{\tiny FN}} h_y^{\mbox{\tiny FN}} +
  \lambda' &=& 0 \; , \quad y > x \; ; \label{NLO_SYSTEM2} \\
  \lambda h_0 + \lambda_{\mbox{\tiny FN}} h_0^{\mbox{\tiny FN}} +
  \lambda' &=& - \mu  \; , \label{NLO_SYSTEM3}
\eea
to be solved for each pair of lattice \textit{distances} $(x,y)$,
$y>x$. The number of independent pairs is $7(7+1)/2 = 28$ for
lattice extension $L = 16$. The solutions, labelled by $x$ and $y$, are 
\bea
  \lambda &=& \frac{h_x^{\mbox{\tiny FN}} - h_y^{\mbox{\tiny
  FN}}}{d_{xy}} \, \mu  \; , \label{L_NLO} \\
  \lambda' &=& \frac{h_x h_y^{\mbox{\tiny FN}} - h_y h_x^{\mbox{\tiny
  FN}}}{d_{xy}} \, \mu  \; , \label{LP_NLO} \\
  \lambda_{\mbox{\tiny FN}} &=& - \frac{h_x - h_y}{d_{xy}} \, \mu  \; ,
  \label{LFN_NLO} 
\eea
where we have defined the determinant
\be
  d_{xy} \equiv h_0 (h_x^{\mbox{\tiny FN}} - h_y^{\mbox{\tiny FN}}) -
  h_0^{\mbox{\tiny FN}} (h_x - h_y) + h_x h_y^{\mbox{\tiny FN}} - h_y
  h_x^{\mbox{\tiny FN}} \; .
\ee
We have checked that the numerical values for $d_{xy}$ are not close to
zero so that there is no problem with small denominators in the
solutions. 
\FIGURE[ht]{\includegraphics[scale=0.52]{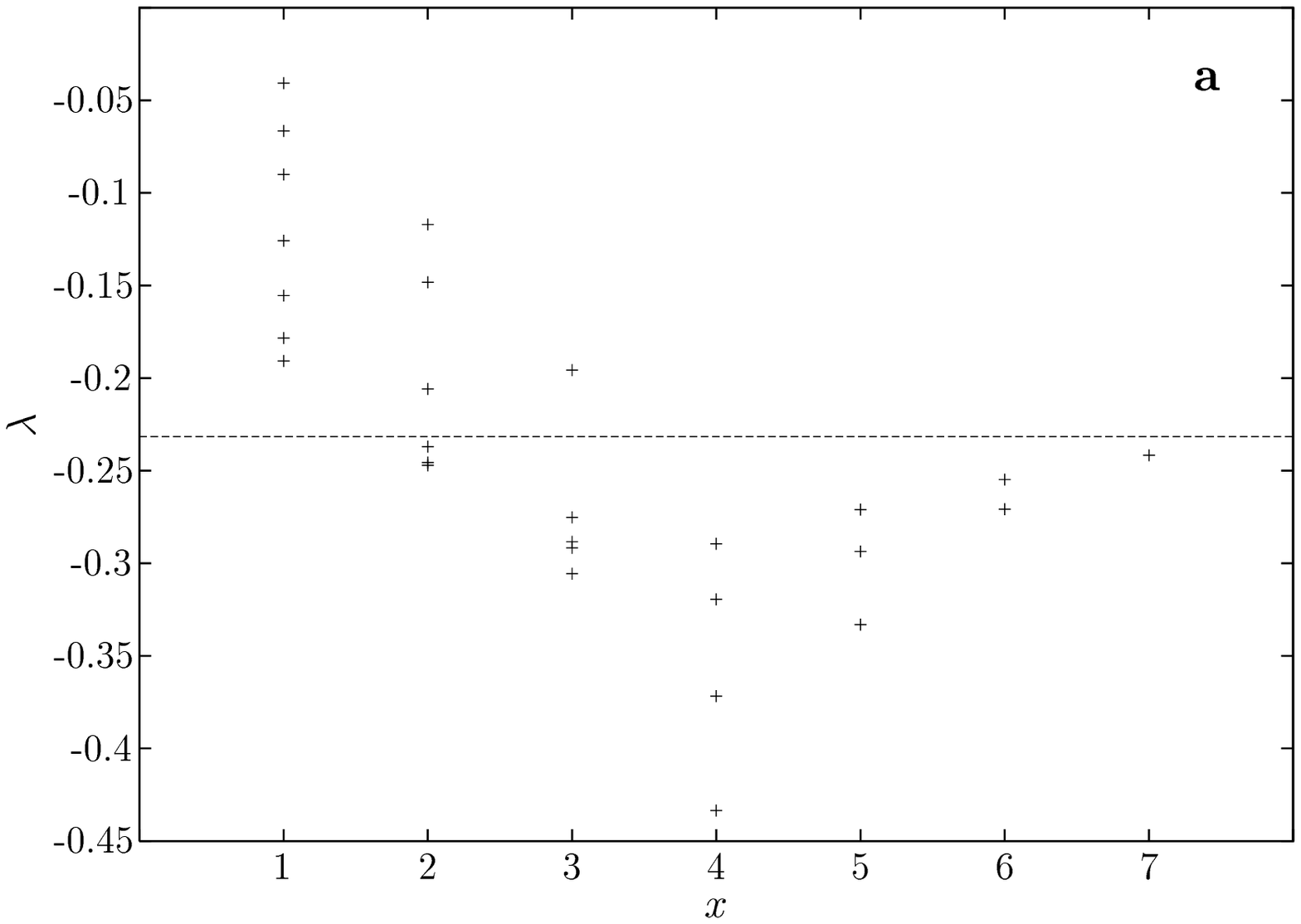}
            \includegraphics[scale=0.52]{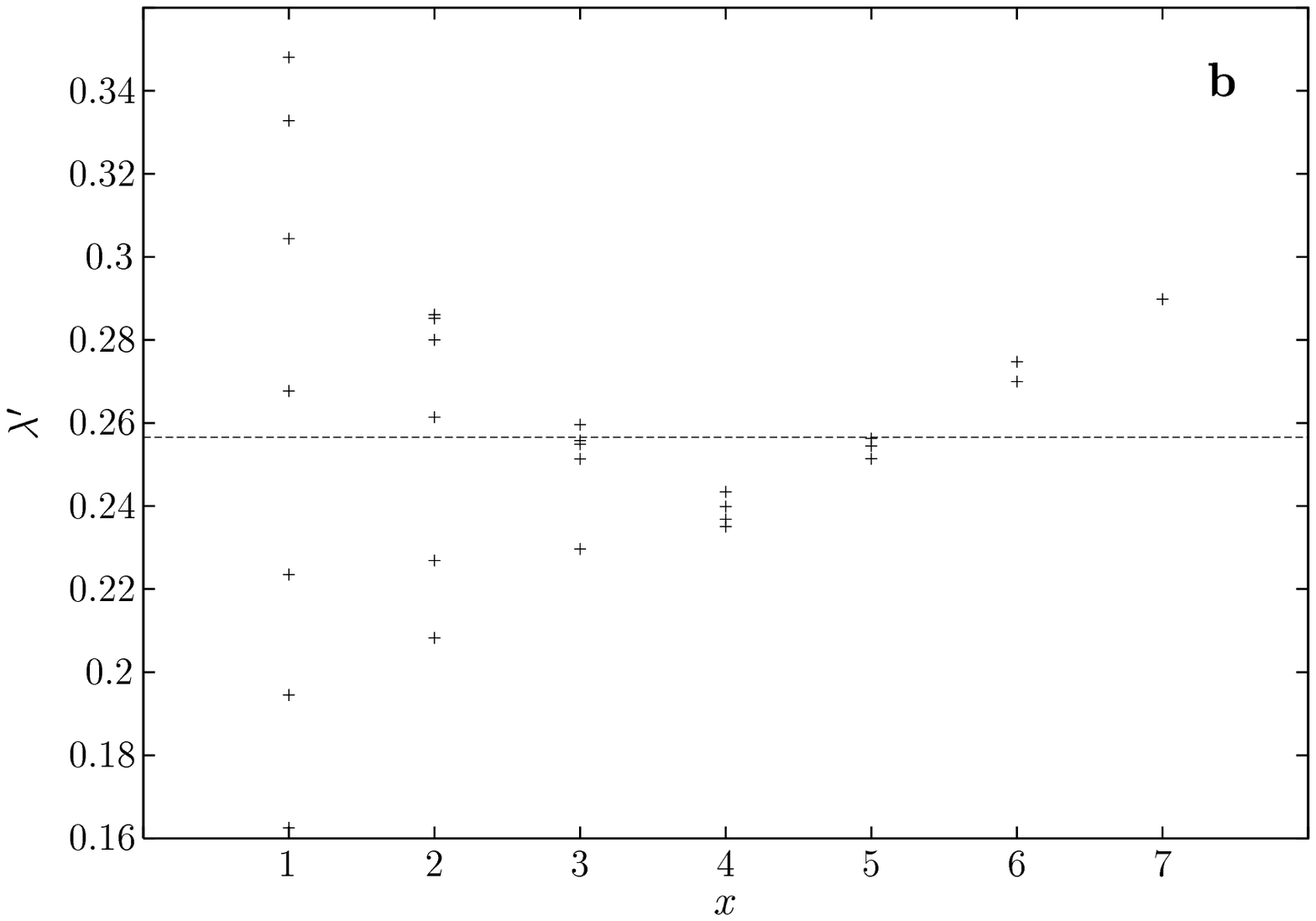}
            \includegraphics[scale=0.52]{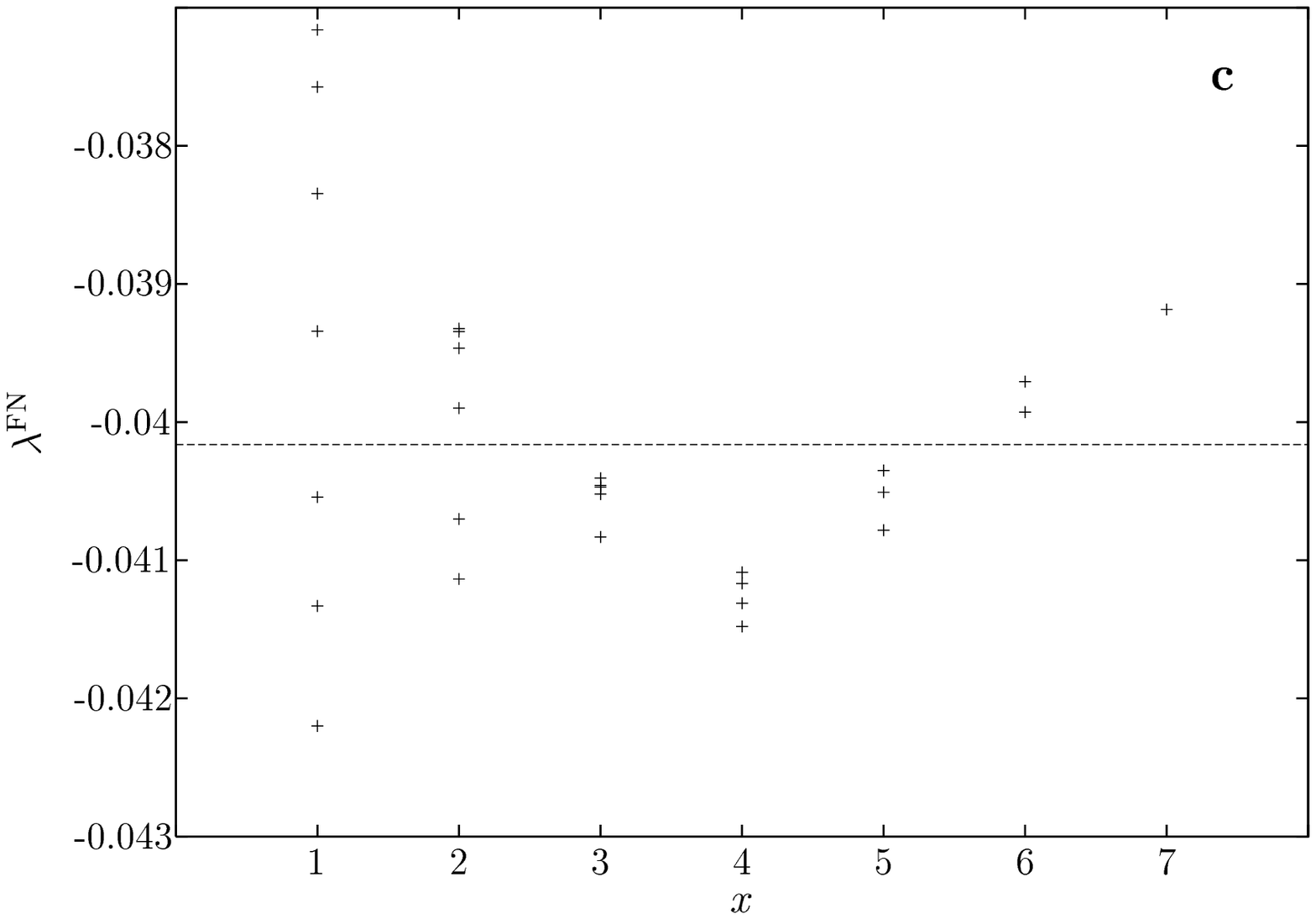} 
\caption{Variation of the couplings
(\protect\ref{L_NLO}--\protect\ref{LFN_NLO}) with lattice distances 
$x$ and $y$. The values associated with $y > x$ are plotted
along  vertical lines labelled by $x = 1 , \ldots , 8$. The horizontal
lines denote the mean value. Note that, for larger distances, the data
seem to deviate less from the central values. \textbf{(a)} coupling $\lambda$,
\textbf{(b)} coupling $\lambda'$, \textbf{(c)} coupling
$\lambda^{\mbox{\tiny FN}}$.}  
\label{FIG:LAMBDAS}}
\clearpage
For each pair of lattice distances $(x,y)$ we thus have a
certain value for any of the three couplings
(\ref{L_NLO}--\ref{LFN_NLO}). For each particular coupling those would all
agree (within statistical errors) if the NLO action (\ref{S_NLO}) would
exactly describe the Yang--Mills ensemble. Again, however, analogous to
the LO case, the couplings do vary with lattice distances $x$ and $y$ as
shown in Fig.~\ref{FIG:LAMBDAS}. Numerically, one finds, 
\bea
  \lambda &=& -0.232 \pm 0.035 \; , \label{L_NUM_NLO} \\
  \lambda' &=& \phantom{-} 0.257 \pm 0.014 \; , \label{LP_NUM_NLO} \\
  \lambda_{\mbox{\tiny FN}} &=& - 0.0402 \pm 0.0004 \label{LFN_NUM_NLO}
  \; .
\eea
Several remarks are in order. First of all, the relative errors, given
by the standard deviation from the mean (see Fig.~\ref{FIG:LAMBDAS}), are
small compared to the LO ansatz. In particular, the signs of all 
couplings are fixed. Interestingly, the addition of the FN coupling
$\lambda_{\mbox{\tiny FN}}$, although small numerically, has a large
effect: it reverts the sign of $\lambda'$ as compared to
(\ref{LP_NUM_LO}), implying a \textit{negative} magnetization. This
follows, for instance, from the Ward identity (\ref{WARD_LO}), which
still holds for the action (\ref{S_NLO}), and the positivity of the
susceptibility, hence
\be
  \mathfrak{M} = - \lambda' \chi^\perp < 0 \; , 
\ee
in contradistinction with the positive Yang--Mills value of Table~1.

To further analyse the result for the couplings, we divide
(\ref{NLO_SYSTEM1}) by $\lambda$,  leading to a linear relation between 
$h$ and $h^{\mbox{\tiny FN}}$ (for $x \ne 0$),  
\be
\label{HVSHFN}
  h_x = - \kappa' - \kappa_{\mbox{\tiny FN}} h_x^{\mbox{\tiny FN}} \; ,
  \quad \kappa_{\mbox{\tiny FN}} \equiv
  \lambda_{\mbox{\tiny FN}}/\lambda \; . 
\ee
Thus, plotting $h_x$ against $h_x^{\mbox{\tiny FN}}$ should yield a
straight line with intercept $-\kappa'$ and slope $-\kappa_{\mbox{\tiny
FN}}$. The numerical values (\ref{L_NUM_NLO}--\ref{LFN_NUM_NLO}) yield
\be
\label{KAPPAS_NUM_NLO}
  \kappa' = - 1.108 \pm 0.228 \; , \quad \kappa_{\mbox{\tiny FN}} = 0.173
  \pm 0.024 \; .
\ee
In analogy with the LO case, we have numerically checked the prediction
(\ref{HVSHFN}) for the NLO action (\ref{S_NLO}) by a Monte Carlo
simulation with 150 configurations using the input couplings
(\ref{L_NUM_NLO}--\ref{LFN_NUM_NLO}). Fig.~\ref{FIG:HVSHFN_FN} clearly
demonstrates the expected linear behavior. A corresponding fit results in  
\be
  \kappa' = - 1.120  \; , \quad \kappa_{\mbox{\tiny FN}} = 0.171  \; ,
\ee
being consistent with the central values of (\ref{KAPPAS_NUM_NLO})
to within one percent.
\FIGURE[ht]{\includegraphics[scale=0.6]{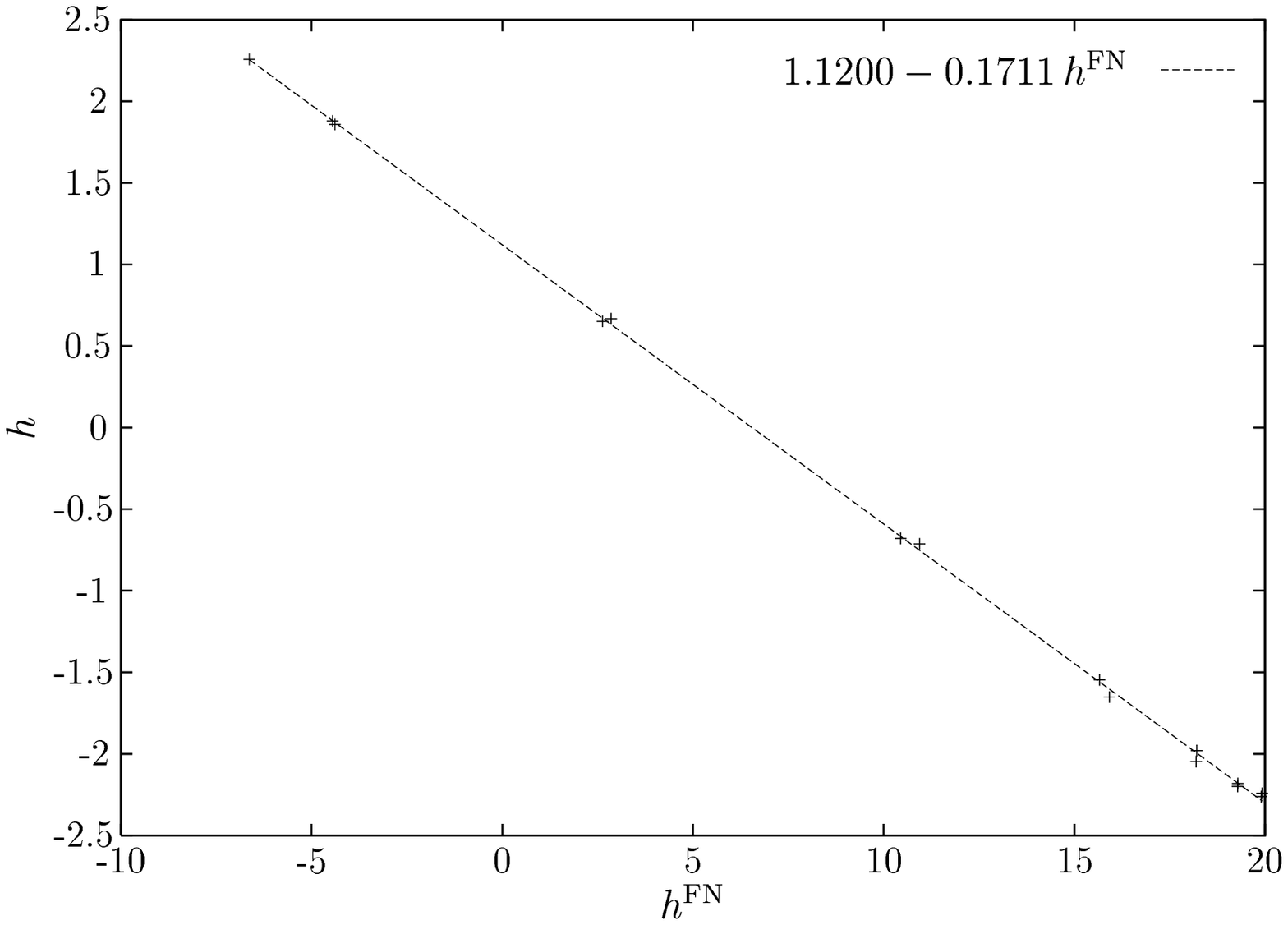} 
\caption{$h$ vs.\ $h^{\mbox{\tiny FN}}$ for the NLO action
(\protect\ref{S_NLO}).}
\label{FIG:HVSHFN_FN}}
We thus conclude that inverse Monte Carlo also works quite well for
the  NLO ensemble. For the sake of explicit comparison with Fig.s 
\ref{FIG:H_YM} and \ref{FIG:H_FN_YM} we display the
reduced two--point functions ob\-tained by simulating the NLO action in
Fig.s \ref{FIG:H_NLO} and \ref{FIG:H_FN_NLO}. 
\FIGURE[ht]{\epsfig{file=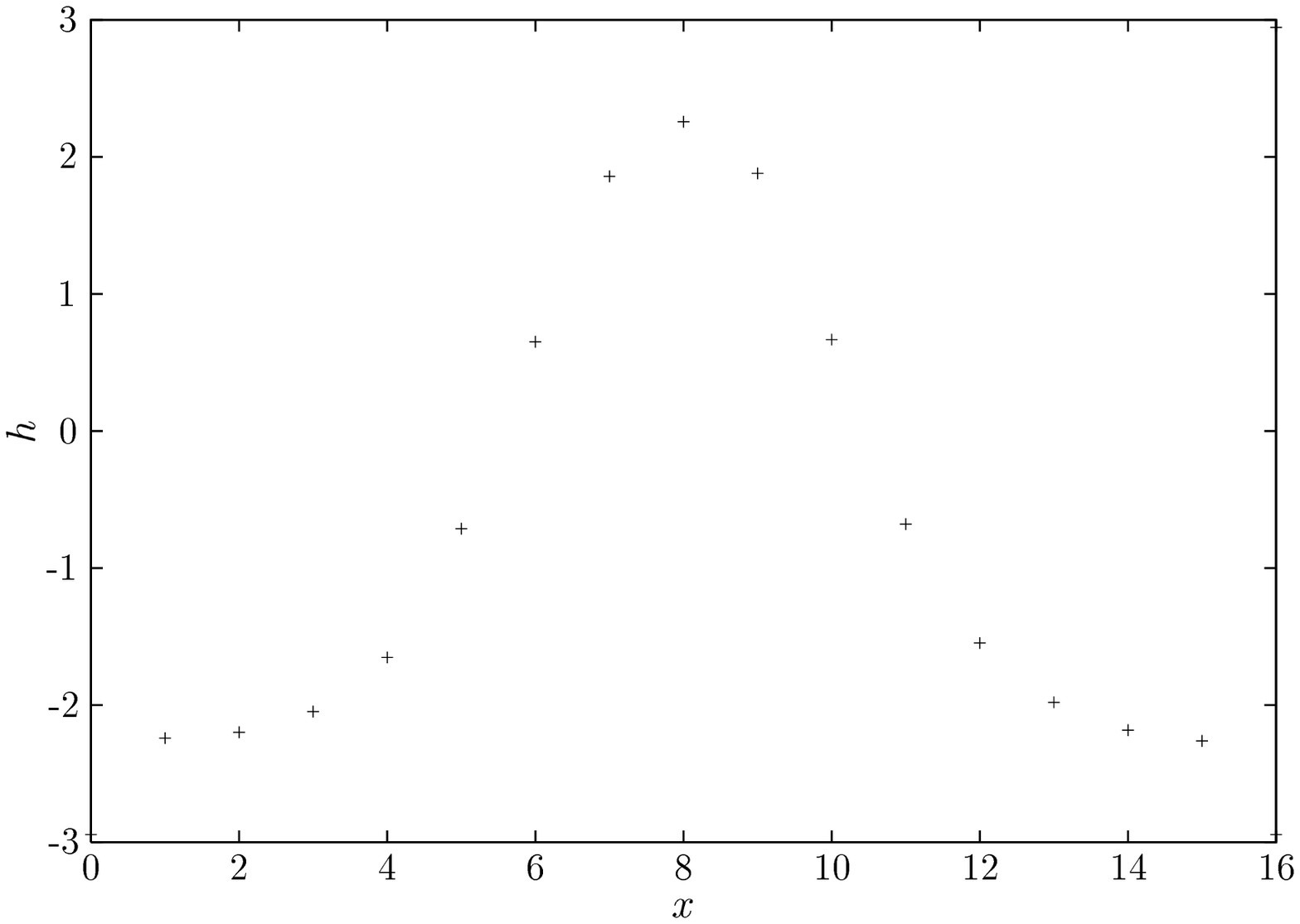,scale=0.6} 
\caption{$h_x$ for the NLO ensemble.}
\label{FIG:H_NLO}}
\FIGURE[ht]{\epsfig{file=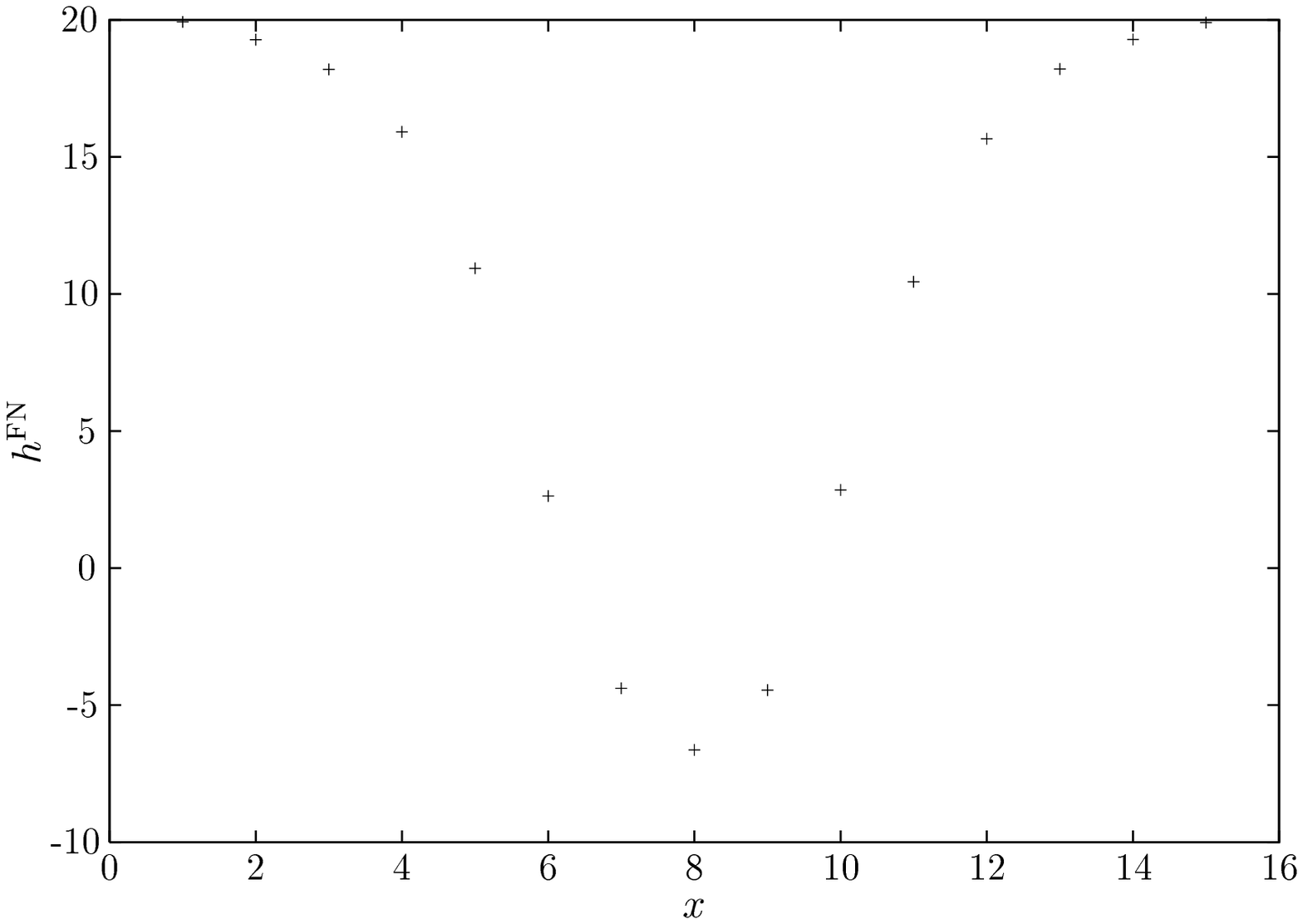,scale=0.6} 
\caption{$h^{\mbox{\tiny FN}}_x$ for the NLO ensemble.}
\label{FIG:H_FN_NLO}}

As expected, the NLO simulation yields a negative magnetization,
\be
  \mathfrak{M} = - 0.49 \; ,
\ee
while the mass gap becomes $M = 1.2$, i.e.~slightly larger than the
value listed in Table~1. 

If the Yang--Mills ensemble has anything to do with the FN one, then
plotting $h$ vs.\ $h^{\mbox{\tiny FN}}$ (as obtained from Yang--Mills)
should also show straight--line behavior, at least
approximately. Fig.~\ref{FIG:HVSHFN_YM} displays a linear fit to the the
Yang--Mills data with parameters
\be
  \kappa' = - 1.232 \; , \quad \kappa_{\mbox{\tiny FN}} = 0.170 \; .
\ee
\FIGURE[ht]{\includegraphics[scale=0.6]{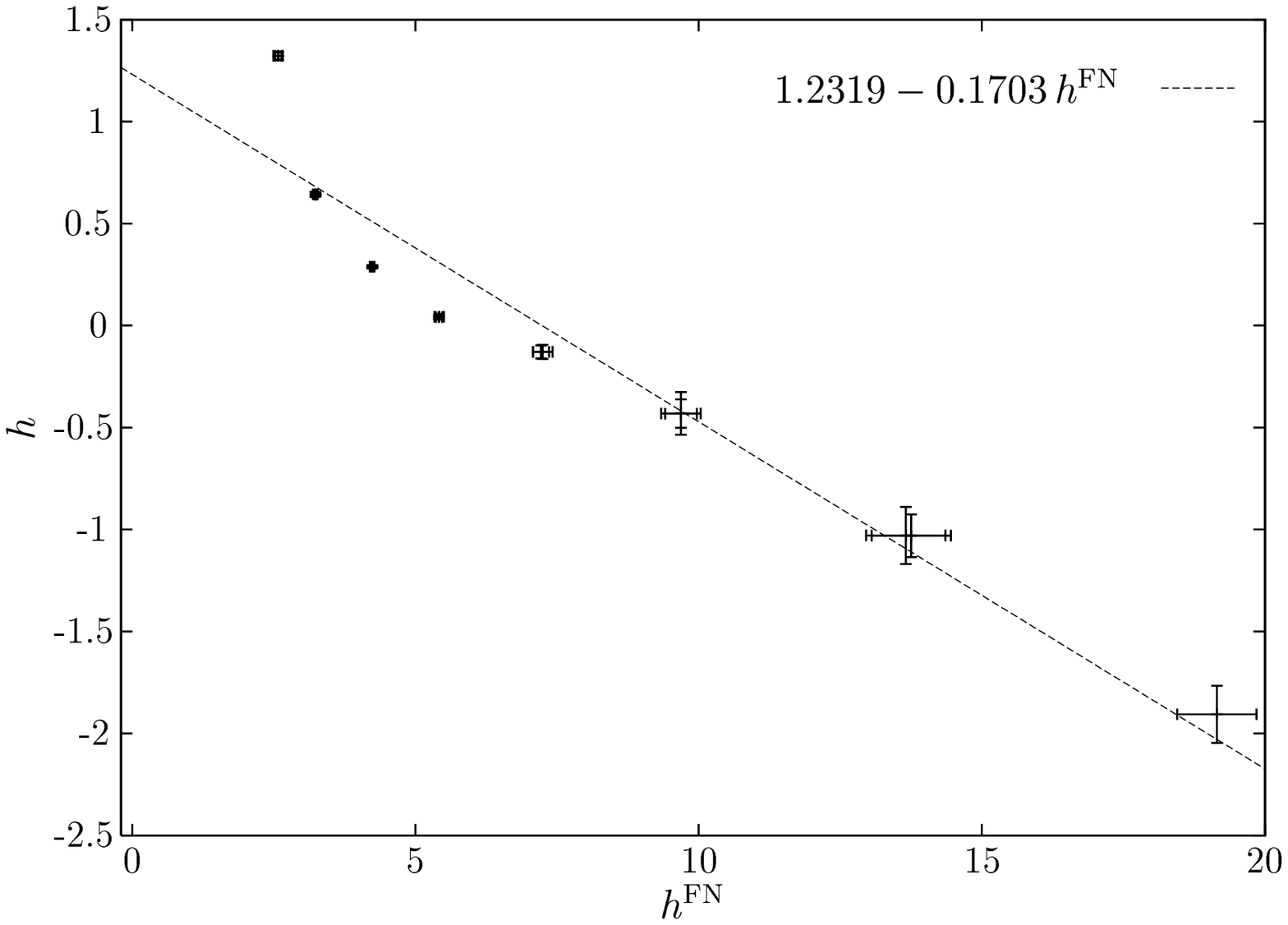} 
\caption{$h$ vs.\ $h^{\mbox{\tiny FN}}$ for the Yang--Mills
ensemble. The error bars reflect the fact that, for large distances
$x$ (i.e.~$x \simeq L/2 = 8$ for the lattice size used), $h$ and
$h^{\mbox{\tiny FN}}$ are obtained by dividing two small numbers (from
the tails of the two--point functions $G^\perp$, $H$ and $H^{\mbox{\tiny
FN}}$).} 
\label{FIG:HVSHFN_YM}}

Again, within error bars, these values are consistent with the preceding
analysis (\ref{KAPPAS_NUM_NLO}). Note that reverting the  sign of
$\lambda'$ amounts to reverting the sign of the intercept in
Fig.~\ref{FIG:HVSHFN_YM}. The data points clearly do not support
anything like that. On the contrary, it seems that, for small
$h^{\mbox{\tiny FN}}$ (corresponding to small distances $x$, see
Fig.~\ref{FIG:H_FN_YM}), the data points deviate from a straight
line. Playing around with different fits indicates that $h$
rises with a \textit{negative} power of $h^{\mbox{\tiny FN}}$ for small
$h^{\mbox{\tiny FN}}$ so that in reality there may be no intercept at
all.  

The behavior of $\kappa'$ as a function of lattice distances $x$ and $y$
may also be investigated. Dividing (\ref{LP_NLO}) by (\ref{L_NLO}) we find
\be
\label{KAPPAPRIME_NUM_NLO}
  \kappa' = \lambda'/\lambda = \frac{h_x h_y^{\mbox{\tiny FN}} - h_y
  h_x^{\mbox{\tiny FN}}}{h_x^{\mbox{\tiny FN}} - h_y^{\mbox{\tiny FN}}}
  = -1.38 \pm 0.29 \; ,
\ee
where, in the analytic expression,  the determinant $d_{xy}$ has dropped out.  
Fig.~\ref{FIG:KAPPA} shows the variation of $\kappa'$ with $x$ and
$y$. Again, a different sign for $\kappa'$ is completely out of reach.  
\FIGURE[ht]{\includegraphics[scale=0.6]{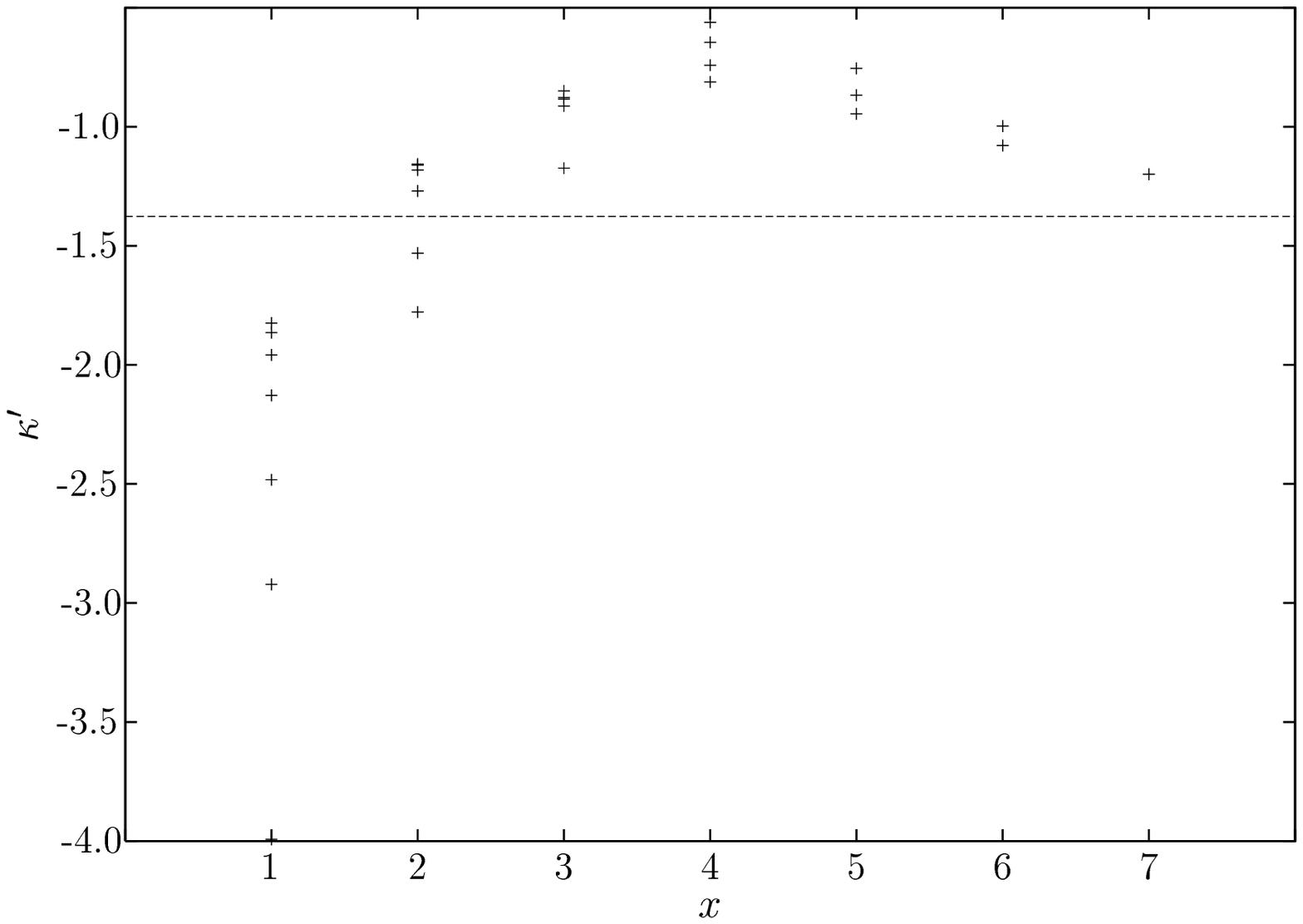} 
\caption{Variation of $\kappa'$ with lattice distances $x$ and $y$. The
central value is $\kappa' = -1.38$.}
\label{FIG:KAPPA}}

Following the logic of gradient expansions, one may argue that the
effective action (\ref{S_NLO}) is supposed to 
represent the Yang--Mills ensemble only for large
distances. Fig.~\ref{FIG:HVSHFN_YM}, for instance, seems to indicate that
the straight--line fit works particularly well for the last three points
to the right which correspond to $x = 6,7,8$, respectively. In physical
units, this amounts to distances $R$ larger than six lattice units,
i.e.\ $R$ $\simgeq$ 0.8 fm. Restricting to the analogous data points in
Fig.~\ref{FIG:LAMBDAS}, we obtain for the couplings in (\ref{S_NLO}), 
\bea
  \lambda &=& - 0.2775 \pm 0.026 \; , \\
  \lambda' &=& \phantom{-} 0.2661 \pm 0.012 \; , \\
  \lambda^{\mbox{\tiny FN}} &=& - 0.040 \pm 0.0005 \; ,
\eea
and for the `reduced' ones,
\be
  \kappa' = -0.97 \pm 0.13 \; , \quad \kappa^{\mbox{\tiny FN}} = 0.15 \pm
  0.01 \; .
\ee
All these do not differ significantly from the values
(\ref{L_NUM_NLO}--\ref{LFN_NUM_NLO}) and (\ref{KAPPAS_NUM_NLO}) obtained
by using the unrestricted data set. In particular, the sign of
$\lambda'$ remains positive. We therefore conclude that, also at large
distances, the minimally
modified FN action (\ref{S_NLO}) fails to describe the Yang--Mills
ensemble of $\vcg{n}$--fields.

\section{Summary and discussion}

We have performed a lattice test of the FN conjecture 
stating that low--energy $SU(2)$ Yang--Mills theory is equivalent to a
Skyrme--type sigma model. More specifically, FN suggest that the knot
solitons of their model might be related to the Yang--Mills glueball
spectrum.  

Using standard Monte Carlo techniques, we have generated an ensemble of
$SU(2)$ link fields from the Wilson action. This ensemble was then used
to extract an associated ensemble of color vectors $\vcg{n}$, with
$\vcg{n}$ parametrizing the gauge invariant distance between the
maximally--Abelian and lattice Landau gauge slices. As these gauges are
close to each other, there is a preferred direction for the
$\vcg{n}$--field which corresponds to \textit{explicit} symmetry breaking. In
this way we avoid the appearance of massless Goldstone bosons and thus
generate a nonvanishing mass gap. A study of the exponential decay of
correlators yields a mass gap close to 1 GeV. To find the effective
action describing the Yang--Mills ensemble of $\vcg{n}$--fields we
have employed inverse Monte Carlo techniques. These are based on
Schwinger--Dyson and Ward identities which we have derived 
analytically on the lattice. The identities have been evaluated
numerically for the Yang--Mills ensemble on the one hand, and for
ensembles stemming from LO and NLO effective actions on the other
hand. As a result, we have found strong evidence that the
ensemble generated from Yang--Mills theory \textit{cannot} be described
by the FN action plus a minimal symmetry--breaking
term to allow for a mass gap. This follows from a number of discrepancies
between the two ensembles. First, and most prominent, the sign of
$\lambda'$ is positive, implying \textit{negative} magnetization $\mathfrak{M}$,
at variance with the value from the 
Yang--Mills ensemble. Second, the reduced two--point function $h$
($h^{\mbox{\tiny FN}}$) from the NLO ensemble \textit{in}creases
(\textit{de}creases) with lattice disctance $x$, while for Yang--Mills
the behavior is just the opposite. Third, the size of the mass gap is
larger than for the Yang--Mills ensemble of $\vcg{n}$--fields.

It is quite conceivable that magnetization (and susceptibility) can be
recovered correctly by adding more (symmetry--breaking) terms to the
NLO action (work in this direction is under way). The same remark
applies to the mass gap. Note, however, that one cannot naturally expect
the Yang--Mills and $\sigma$--model mass gaps to coincide due
to the nonlocal relation between $\vcg{n}$ and the link variables
$U$\footnote{ As there is no unique or natural definition for $\vcg{n}$, one
may try alternative prescriptions for $\vcg{n} = \vcg{n}[U]$. A fairly
local one is 
the following. Write the (gauge fixed) links as $U_{x, \mu} = u_{x,
\mu}^0 + i u_{x, \mu}$. Then define $\bar{n}_x \equiv \bar{u}_x /
|\bar{u}_x|$ with the link average $\bar{u}_x \equiv \sum_\mu u_{x,
\mu}$. Under global gauge transformations this transforms properly such
that $\bar{\vcg{n}}$ is another color unit vector.}. Whether this
represents a problem is a question of scales. If the effective
$\sigma$--model were valid only for distances of, say, $R$ $\simgeq$ 0.8
fm corresponding to energies $E$ $\simleq$ 0.25 GeV, as suggested by the
discussion of Section~5, then it would make
no sense to address questions like the glueball
spectrum. An analogous situation holds for the Fermi theory of
weak interactions which also is only effective much below the $W$ and
$Z$ scales.

Finally, one should mention that there is still another fundamental
problem associated with describing Yang--Mills theory in terms of
effective $\sigma$--models. Allowing for finite temperature, the latter
are in the universality class of the 4$d$ Heisenberg model, while 
$SU(2)$ Yang--Mills theory is in the 3$d$ Ising class
\cite{svetitsky:82}. This issue has been discussed recently in the context
of constructing effective actions via Abelian projections
\cite{yee:95,ogilvie:99,ogilvie:02}. Again, if the  $\sigma$--model
scale were below the critical temperature, the effective theory would
only be valid in the confined phase and would have nothing to say about
the behavior of Yang--Mills theory close to the phase
transition. Otherwise, an infinite 
number of operators would be required which, of course, is anything else
but an `effective' description. Summarizing, we conclude that,  while a
reasonable effective model generalizing the FN action may exist in
principle, it will  be of little practical use.

\acknowledgments

The authors are indebted to S.~Shabanov for suggesting this
investigation and to P.~van Baal for raising the issue of Goldstone
bosons. Discussions with F.~Bruckmann, P.~de~Forcrand, A.~di
Giacomo, A.~Gonz{\'a}lez-Arroyo, D.~Hansen, M.~Hasenbusch, K.~Langfeld,
E.~Seiler, M.~Teper, and V.~Zakharov are gratefully acknowledged.
The work of T.H.\ was supported by DFG under contract Wi 777/5-1. He
thanks the theory department of LMU Munich, in particular V.~Mukhanov,
for the hospitality extended to him. L.D.\ thanks G.~Bali for providing
the code for algorithm AI as well as helpful advice.

\appendix

\section{Conventions}
\label{APP:CONVENTIONS}

Left and right lattice derivatives are defined as 
\bea
  \partial_\mu f_x &\equiv& f_{x + \mu} - f_x \; , \\
  \partial_\mu^\dagger  f_x &\equiv& f_{x - \mu} - f_x \; .
\eea
The ordinary lattice Laplacian $\triangle \equiv -
\partial_\mu^\dagger \partial_\mu$ is  a negative semi--definite
operator. Its action on lattice functions $f$ is given by
\be
  \triangle f_x \equiv - \sum_\mu (2f_x - f_{x + \mu} - f_{x - \mu}) \;
  .
\ee
The \textit{covariant} Laplacian $\triangle [U]$ in the adjoint
representation acts as
\be
  \triangle^{ab}[U] f_x^b \equiv -\sum_\mu \left( 2 f_x^a - R_{x,
  \mu}^{ab} \, f_{x + \mu}^b - R_{x- \mu, \mu}^{ba} \, f_{x - \mu}^b \right) 
\; , 
\ee
where we have defined the \textit{adjoint link}
\be
\label{ADJ_LINK}
  R_{x,\mu}^{ab} \equiv \frac{1}{2} \tr (\tau^a U_{x,\mu} \tau^b
  U_{x,\mu}^\dagger) \; .
\ee

\section{Relating LLG and MAG}
\label{APP:LG_MAG}

From (\ref{F_LG}) it follows immediately that the LLG
\textit{minimizes} the functional \cite{faber:01} 
\be
  \bar{F}_{\mathrm{LLG}} \equiv \sum_l \tr (\Eins - {}^\Omega U_l) \; ,
  \quad  l \equiv (x , \mu) \; ,    
\ee
and thus tends to  bring the links $U_l$ close to $\Eins$. The MAG, on 
the other hand, minimizes 
\be
   \bar{F}_{\mathrm{MAG}} \equiv \sum_l  (1 - {}^g R_l^{33}) \; , 
\ee
and thus wants to bring the 33--entry of the adjoint link $R_l^{ab}$
close to 1. From (\ref{ADJ_LINK}) it is obvious that, if $U_l$ equals
unity, the same will be true for $R_l$.  This can be made more
precise: if $U_l \simeq \Eins + i a A_l$, $A_l$ hermitean, then it is an
easy exercise to show that $R_l^{33} = 1 + O(a^2)$. In this sense, 
the LLG is close to the MAG.

\section{Schwinger--Dyson equations and Ward identities}
\label{APP:WARD}

We begin with computing the infinitesimal rotations of the various
contributions in (\ref{SYMM}) and (\ref{NONSYMM}) to the effective
action. It turns out that, for all $S_j$,  the action of the angular
momentum can be written as
\be
  i \vcg{L}_x S_j = \vcg{n}_x \times \vcg{K}_{jx} \; ,
\ee
(and analogous for the $S_k'$) with the vectors $\vcg{K}_{jx}$  and
$\vcg{K}_{kx}'$ given by 
\begin{eqnarray}
\label{identity4}
  \vcg{K}_{1x} &=& 2 \triangle \vcg{n}_x  \\ 
  \vcg{K}_{2x} &=& 2 \triangle^2 \, \vcg{n}_x  \\ 
  \vcg{K}_{3x} &=& 2 \left[ \triangle 
  \vcg{n}_x (\vcg{n}_x \cdot \triangle \vcg{n}_x) + \triangle \left(\vcg{n}_x
  (\vcg{n}_x \cdot \triangle \vcg{n}_x) \right) \right] \label{LS3} \\ 
  \vcg{K}_{4x} &=& 2 \left[ \pa_\mu^\dagger \pa_\nu
  \vcg{n}_x (\vcg{n}_x \cdot  \pa_\mu^\dagger \pa_\nu \vcg{n}_x) +
  \pa_\mu^\dagger \pa_\nu  \left(\vcg{n}_x 
  (\vcg{n}_x \cdot \pa_\mu^\dagger \pa_\nu \vcg{n}_x) \right) \right]
  \label{LS4} \\ 
  \vcg{K}_{1x}' &=& \vcg{h}  \\ 
  \vcg{K}_{2x}' &=& 2 \vcg{h} \, (\vcg{n}_x \cdot
  \vcg{h})    \\
  \vcg{K}_{3x}' &=&  \vcg{h} \, (\vcg{n}_x \cdot
  \triangle \vcg{n}_x)  + \triangle \vcg{n}_x (\vcg{n}_x \cdot \vcg{h})
  + \triangle   (\vcg{n}_x (\vcg{n}_x \cdot \vcg{h}))   \; .
\end{eqnarray}
Choosing the $F$'s in (\ref{WARD}) as $n_x^a$, $n_x^a n_y^b$ and $n_x^a
n_y^b n_z^c$, respectively, results in the Ward identities
\bea
  G_{\underline{x}y}^{ii} \lambda_1' + 2
  G_{\underline{x}\underline{x}y}^{3ii} \lambda_2' +
  G_{\underline{x}\underline{x}y}^{\Delta ii} \lambda_3' &=& -2 G_y^3
  \label{GEN_WARD1} \; , \\
  G_{\underline{x}yz}^{ii3} \lambda_1' + 2
  G_{\underline{x}\underline{x}yz}^{3ii3} \lambda_2' +
  G_{\underline{x}\underline{x}yz}^{\Delta ii3} \lambda_3' &=& -2
  G_{yz}^{33} + G_{yz}^{ii}  \label{GEN_WARD2} \; , \\
  G_{\underline{x}yzz'}^{ii33} \lambda_1' + 2
  G_{\underline{x}\underline{x}yzz'}^{3ii33} \lambda_2' +
  G_{\underline{x}\underline{x}yzz'}^{\Delta ii33} \lambda_3' &=& -2
  G_{yzz'}^{333} + 2 G_{yzz'}^{i(i3)} \label{GEN_WARD3} \; .
\eea
Here, the superscript $(i3)$ denotes symmetrization in $i$, 3, and we
have introduced the shorthand notations 
\bea
  G_{xyz \ldots}^{abc \ldots} &\equiv& \bra n_{x}^{a}
  n_{y}^b n_z^c \ldots \ket \; ,  \\
  G_{\underline{x}y \ldots}^{ii \ldots} &\equiv& \sum_x
  G_{xy\ldots}^{ii\ldots} \; , \\
  G_{\underline{x}\underline{x}y\ldots}^{3ii\ldots} &\equiv& \sum_x \bra
  n_x^3 n_x^i n_y^i \ldots \ket \; ,  \\
  G_{\underline{x}\underline{x}y\ldots}^{\Delta ii\ldots} &\equiv&
  \sum_x \bra (n_x^a \triangle n_x^a) n_x^i n_y^i \ldots \ket \; , 
\eea
with $\underline{x}$ denoting summation over all lattice sites $x$. In
particular, one has
\be
\label{PARTICULAR}
  G_y^3 \equiv \mathfrak{M} \; , \quad G_{\underline{x}y}^{ii} \equiv 2
  \chi^\perp \; , \quad G_{xy}^{ii} \equiv 2 G_{xy}^\perp \; , \quad
  \quad G_{xy}^{33} \equiv G_{xy}^\parallel \; .
\ee
It seems that a particular recurrence pattern arises in
(\ref{GEN_WARD1}--\ref{GEN_WARD3}) 
that could be used to derive a Ward identity for an insertion of an
arbitrary number of $\vcg{n}$'s. For the NLO  derivative
expansion, however, the three identities are sufficient to determine the
symmetry--breaking couplings $\lambda'$. 

For the particular case that $F$ in (\ref{SD}) equals $n$, $F_x^a =
n_x^a$, we can give the general Schwinger--Dyson equation in closed
form,
\be
  \sum_j \lambda_j H_{j,xy} + \sum_k \lambda_k' H_{k,xy}' = -
  \mathfrak{M} \delta_{xy} \; , 
\ee
where we have introduced the two--point function
\be
  H_{j,xy} \equiv \bra n_x^i n_y^{[\, i} K_{jy}^{3]} \ket \; ,
\ee
and, completely analogous, $H_{k,xy}'$.  The FN terms are (\ref{LS3})
minus (\ref{LS4}), so that the relevant two--point function becomes
\be
\label{H^FN} 
  H_{xy}^{\mbox{\tiny FN}} \equiv H_{3,xy} - H_{4,xy}   \; , 
\ee
which has been used in (\ref{SDE_NLO}).

\bibliographystyle{JHEP.bst}
\bibliography{../../bibfiles/gauge}

\end{document}